\documentclass[twocolumn,showpacs,preprintnumbers,amsmath,amssymb]{revtex4}
\usepackage{amsmath,amsfonts,latexsym,amssymb,graphics,epsfig,subfigure,color,makeidx,inputenc}
\usepackage{xcolor}
\usepackage{graphicx}
\usepackage{hyperref}

\def\gw#1{gravitational wave#1 (GW#1)\gdef\gw{GW}}
\def\bbh#1{binary black hole#1 (BBH#1)\gdef\bbh{BBH}}
\def\bh#1{black hole#1 (BH#1)\gdef\bh{BH}}
\def\maya#1{\textsc{Maya}#1}
\def\et#1{\textsc{EinsteinToolkit}#1}
\def\twopunctures#1{\textsc{2Punctures}#1}

\graphicspath{{./Figs/}}

\begin{document}

\title{Gravitational recoil from binary black hole mergers in scalar field clouds}

\author{Yu-Peng Zhang$^{a}$, 
		Miguel Gracia-Linares$^{b}$, 
        Pablo Laguna$^{b}$, 
        Deirdre Shoemaker$^{b}$, 
        Yu-Xiao Liu$^{a}$
}
\affiliation
{
$^a$Lanzhou Center for Theoretical Physics, Key Laboratory of Theoretical Physics of Gansu Province, Institute of Theoretical Physics, Research Center of Gravitation, School of Physical Science and Technology, Lanzhou University, Lanzhou 730000, China\\
$^b$Center of Gravitational Physics, Department of Physics, University of Texas at Austin, Austin, TX 78712, U.S.A.}

\begin{abstract}
In vacuum, the gravitational recoil of the final black hole from the merger of two black holes depends exclusively on the mass ratio and spins of the coalescing black holes, and on the eccentricity of the binary. If matter is present, accretion by the merging black holes may modify significantly their masses and spins, altering both the dynamics of the binary and the gravitational recoil of the remnant black hole. This paper considers such scenario. We investigate the effects on the kick of the final black hole from immersing the binary in a scalar field cloud. We consider two types of configurations: one with non-spinning and unequal-mass black holes, and a second with equal mass and spinning holes. For both types, we investigate how the gravitational recoil of the final black hole changes as we vary the energy density of the scalar field. We find that the accretion of scalar field by the merging black holes could have a profound effect. For the non-spinning, unequal-mass binary black holes, the kicks are in general larger than in the vacuum case, with speeds of $\sim 1,200~\text{km/s}$ for binaries with mass ratio 2:1, one order of magnitude larger than in vacuum. For equal mass, binaries with black holes with spins aligned with the orbital angular momentum, kicks larger than in vacuum are also found. For systems with spins in the super-kick configuration, the scalar field triggers a similar dependence of the kicks with the entrance angle at merger as in the vacuum case but in this case depending on the strength of the scalar field. 
\end{abstract}

\maketitle

\section{Introduction}\label{sec:intro}

The \gw{s} emitted during the inspiral and coalescence of a \bbh{} carry energy, angular momentum, and linear momentum  \cite{Ruiz_2007}. A net loss of linear momentum by the binary in certain direction implies a recoil of the final \bh{} in the opposite direction \cite{Fitchett1983,Sperhake2007,2009Healy}. In vacuum, this recoil or kick depends exclusively on the mass ratio and spins of the coalescing \bh{}s, and if the binary is not in a quasi-circular orbit, the recoil  depends also on the eccentricity of the binary system~\cite{2007ApJ...656L...9S}. When matter is present, the situation is more complex. For instance, in mixed binary mergers, i.e. coalesences of \bh{s} with neutron stars, the kick will depend also on any  accretion of matter by the \bh{} during the merger~\cite{2021CQGra..38r5008K}. 

For this work, we focus on \bh{} environments permeated by a scalar field. Scalar fields have been considered as sources of dark matter \cite{Marsh:2015xka}, in inflationary theories \cite{Bezrukov:2007,Burgess2009,Burgess2010,Giudice2011,Lerner2010,Lyth1999}, and in the context of modified theories of gravity, such as scalar-tensor  and $f(R)$ theories~\cite{Wagoner1970,Felice2010,Sotiriou2010}. 
In the presence of \bh{}s, scalar fields have also been used to probe the transition from  inspiraling  \bh{s} to a single perturbed \bh{} \cite{Bentivegna2008}. \bbh{} systems in scalar-tensor~\cite{Healy:2011ef,Berti:2013gfa}, $f(R)$~\cite{Cao:2013osa},  and Einstein-Maxwell-dilation~\cite{Hirschmann:2017psw} theories have been also studied, as well as \bbh{s} in dynamical Chern-Simons gravity~\cite{Okounkova:2017yby}, axion-like scalar fields~\cite{Yang:2017lpm}, and scalar Gauss-Bonnet gravity \cite{Witek:2018dmd}. 

Here, we are interested in investigating the effect that a scalar field may have on the kick of the final \bh{}, an aspect not considered by the studies mentioned above. We focus on a simple scenario, a \bbh{} immersed in a spherical shell of a massive scalar field and study two types of \bbh{} configurations. One consists of un-equal mass binaries with non-spinning \bh{s}, and in the other, binaries with equal-mass holes but spinning \bh{s}. For the later, we consider \bh{} spins aligned with the orbital angular momentum (i.e., non-precessing binaries) and \bh{} spins in the orbital plane in the super-kick configuration~\cite{PhysRevLett.98.231101,PhysRevLett.98.231102}. In addition to the kick on the final \bh{}, we also studied the characteristics of the \gw{s} and the angular momentum radiated in \gw{s} and by the scalar field. 

The paper is organized as follows. In Sec.~\ref{sec:initial}, we present the method to construct initial data. Sec.~\ref{sec:evolution} summarizes the equations of motion for the \bbh{} with scalar field sources. Sec.~\ref{sec:extraction} presents the methodology to extract kicks, energy, and angular momentum radiated. The \bbh{} configurations are given in Sec.~\ref{sec:results}. Results for un-equal mass, non-spinning \bh{s} binaries are given in Sec.~\ref{sec:results1} and for equal mass, spinning \bh{s} binaries in Sec.~\ref{sec:results2}. Conclusions are found in Sec.~\ref{sec:conclusions}. Greek indices denote space-time indices, and Latin indices are used for spatial indices.  We use geometrical units in which $G=c=1$. A subscript 0 denotes initial values. Unless explicitly stated, we report results in units of $M_0$, the total initial mass of the \bbh{} system.

\section{Initial Data}\label{sec:initial}
Under a 3+1 decomposition of the Einstein field equations~\cite{baumgarte_shapiro_2010}, the initial data consist of ($\gamma_{ij}, K_{ij}, \rho, S_i$), with $\gamma_{ij}$ the spatial metric and $K_{ij}$ the extrinsic curvature of the constant time, space-like hypersurfaces. $\rho$ and $S_i$ are the energy and momentum densities, respectively. The initial data must satisfy the following equations:
\begin{eqnarray}
R+K^2-K_{ij}K^{ij}&=& 16\pi\rho\label{eq:Ham} \\
\nabla_jK^j_i-\nabla_iK &=& 8\pi S_i\,,\label{eq:mom}
\end{eqnarray}
namely  the Hamiltonian and momentum constraints, respectively. Here $R$ is the Ricci scalar, and $\nabla$ denotes covariant differentiation associated with $\gamma_{ij}$. For our case of a massive scalar field:
\begin{eqnarray}
\rho &=& \frac{1}{2}\Pi^2+\frac{1}{2}\nabla^i\nabla_i\phi+\frac{1}{2}m_\phi^2\phi^2,\label{energydesity}\\
S_i    &=& -\Pi\,\partial_i\phi,\label{energyflux}
\end{eqnarray}
with  $m_\phi$ the mass of the scalar field $\phi$ and $\Pi$ its conjugate momentum. 

We solve the constraints (\ref{eq:Ham}) and (\ref{eq:mom}) following the York-Lichnerowicz conformal approach \cite{Lichnerowicz1944,York1971,York1972,Cook2000} in which
\begin{eqnarray}
\gamma_{ij}&=&\psi^4\eta_{ij}\\
K_{ij} &=&A_{ij} = \psi^{-2}\tilde{A}_{ij}\,,
\end{eqnarray}
with $A^i\,_i=0$, $K = 0$, and $\eta_{ij}$ the flat metric.
In addition, we impose $\phi = \tilde\phi$ and  $\Pi=\psi^{-6}\widetilde{\Pi}$~\cite{Laguna1991,Balakrishna2006}. With these  transformations, the Hamiltonian \eqref{eq:Ham} and the momentum \eqref{eq:mom} constraints read respectively:
\begin{eqnarray}
\Delta\psi+\frac{1}{8}\tilde{A}^{ij}\tilde{A}_{ij}\psi^{-7} &=& -\pi\widetilde{\Pi}^2\psi^{-7} -\pi\psi\partial^i\phi\,\partial_i\phi\nonumber\\
&-& \pi m_\phi^2\phi^2\psi^5\label{hamconstraint}\\
\partial_j\tilde{A}^j_i &=& - 8\pi\widetilde{\Pi}\partial_i\phi\,,\label{momconstraint}
\end{eqnarray}
where $\Delta = \eta^{ij}\partial_i\partial_j$.

Since we are modeling \bh{s} as punctures, the conformal factor $\psi$ diverges at the punctures. Therefore, we will exploit the freedom for choosing initial data for $\phi$ and $\Pi$ and zero out the divergent terms proportional to $\psi$ and $\psi^5$ in Eq.~(\ref{hamconstraint}). We accomplish this by setting initially $\phi=0$. With this assumption, (\ref{hamconstraint}) and (\ref{momconstraint}) become
\begin{eqnarray}
\Delta\psi+\left(\frac{1}{8}\tilde{A}^{ij}\tilde{A}_{ij}+\pi\widetilde{\Pi}^2\right)\psi^{-7}&=&0 \label{hamconstraint2}\\
\partial_j\tilde{A}^j_i &=& 0\,,\label{momconstraint2}
\end{eqnarray}
respectively.

In Eq.~(\ref{momconstraint}), we use the Bowen-York solutions for $\tilde A_{ij}$~\cite{1980PRDBowen}. Since we are interested in asymptotically flat solutions to the conformal factor, we require $\widetilde\Pi$ to have compact support. For simplicity, we set 
\begin{equation}
\widetilde{\Pi}(r)=\Pi_0 \exp\left[-\frac{1}{2}\left(\frac{r-r_0}{\sigma}\right)^2\right]\,.\label{momentum}
\end{equation}
That is, the scalar field source is a shell with radius $r_0$, thickness $\sigma$, and amplitude $\Pi_0$. We solve Eq.~(\ref{hamconstraint2}) equation with the \twopunctures{} solver~\cite{Ansorg2004}, which was modified to include the $\widetilde\Pi^2$ term.

\section{Evolution Equations}\label{sec:evolution}

The evolution equation for the scalar field is
\begin{equation}
\square\phi = m_\phi^2\phi,\label{equationofscalarfield}
\end{equation}
with $\square = \nabla^\mu\nabla_\mu$ and $\nabla_\mu$  covariant differentiation with respect to the space-time metric $g_{\mu\nu}$. Under a 3+1 decomposition, the space-time metric is decomposed as
\begin{equation}
g_{\mu\nu}=\gamma_{\mu\nu}-n_\mu n_\nu,\label{eq:metric}
\end{equation}
with  $n^{\mu}=(\alpha^{-1},-\beta^i \alpha^{-1})$  the time-like unit normal vector to the $t=$ constant space-like hypersurfaces. Here $\alpha$ and $\beta^i$ are the lapse function and shift vector, respectively. Given (\ref{eq:metric}), we rewrite Eq.~(\ref{equationofscalarfield}) as 
\begin{eqnarray}
\frac{1}{\alpha}\partial_o\phi&=&-\Pi,\\
\frac{1}{\alpha}\partial_o\Pi&=&  -\nabla^i\nabla_i\phi -\nabla_i\ln\alpha \nabla^i\phi+  K\Pi+ m_\phi^2\phi\,,
\end{eqnarray}
where $\partial_o = \partial_t-\beta^i\partial_i$.

The evolution of the geometry of the space-like hypersufaces, namely $\gamma_{ij}$ and $K_{ij}$, is handled with the BSSN formulation of the Einstein equations \cite{Shapiro1999,Shibata1995}. For a scalar field, the stress-energy tensor source in these equations is given by 
\begin{equation}
S_{ij} = \nabla_i\phi\nabla_j\phi + \frac{1}{2}\gamma_{ij}(\Pi^2-\nabla^k\nabla_k\phi
-m_\phi^2\phi^2)\,.
\end{equation}
We used the moving puncture gauge~\cite{Campanelli2005,Baker2006} to evolve $\alpha$ and $\beta^i$.
The resulting set of evolution equations is solved numerically using the \maya{} code~\cite{2003VPPR5ICGoodale,2006CPCHusa,2012ApJHaas,2015ApJLEvans,2016PRDClark,2016CQGJani}, our local version of the \et{} code~\cite{EinsteinToolkit:2021_11}.

\section{Physics Extraction}\label{sec:extraction}
The physical quantities of interest are the spin and masses of the \bh{s}, as well as the properties of the radiated emission. The \bh{} masses and spins are computed using the dynamical apparent horizons framework~\cite{2004LRR.....7...10A} as implemented in the \et{}~\cite{EinsteinToolkit:2021_11}.
On the other hand, the energy, linear and angular momentum radiated are computed from the Weyl scalar $\Psi_4$  as follows~\cite{Ruiz_2007}:

\begin{eqnarray}
\frac{dE^{\text{gw}}}{dt}&=&\lim_{r\to\infty}\frac{r^2}{16\pi}\oint \left|\int_{-\infty}^{t}\Psi_4dt'\right|^2d\Omega,\label{rediated-energy}\\
\frac{dP_i^{\text{gw}}}{dt}&=&\lim_{r\to\infty}\frac{r^2}{16\pi}\oint \hat{l}_i\left|\int_{-\infty}^{t}\Psi_4dt'\right|^2d\Omega,
\label{rediated-momentum}\\
\frac{dJ_i^{\text{gw}}}{dt}&=&-\lim_{r\to\infty}\frac{r^2}{16\pi}\text{Re}\bigg[\oint\left(\int_{-\infty}^{t}\bar\Psi_4dt'\right)\nonumber\\
&\times&\hat J_i\left(\int_{-\infty}^{t}\int_{-\infty}^{t'}\Psi_4dt''dt'\right)d\Omega\bigg]\,,
\label{rediated-angular-momentum}
\end{eqnarray}
where  $d\Omega=\sin\theta d\theta d\varphi$, $\hat{l}_i=(\sin\theta\cos\varphi, \sin\theta\sin\varphi,\cos\theta)$, and $\hat J_i$ is the angular momentum operator. Integration of (\ref{rediated-momentum}) yields the recoil or kick of the final \bh{} from the emission of \gw{s}.

In addition to \gw{} emission, we also have emission of energy, linear, and angular momentum associated with the scalar field. We compute this emission  following the method in Ref.~\cite{Witek:2018dmd} as follows:
\begin{eqnarray}
\frac{dE^{\text{sf}}}{dt}&=&\lim_{r\to \infty}r^2\oint{T_{tr}\,d\Omega},\\
\frac{dP_i^{\text{sf}}}{dt}&=&\lim_{r\to \infty}r^2\oint{ T_{ir}\,d\Omega},\\
\label{rediated-momentum-sf}
\frac{dJ_z^{\text{sf}}}{dt}&=&\lim_{r\to \infty}r^2\oint{T_{\phi r}\,d\Omega},
\end{eqnarray}
where the components of the stress-energy tensor are given by
\begin{equation}
T_{\mu\nu} = \nabla_\mu\phi\,\nabla_\nu\phi-g_{\mu\nu}\bigg(\frac{1}{2}\nabla_\alpha\phi\nabla^\alpha\phi+\frac{1}{2}m_\phi^2\phi^2\bigg).\label{eq:Tmunu}
\end{equation}

In all these fluxes, we evaluate the integrals at a finite radius and then extrapolate the values to infinity.

\section{Binary Configurations}\label{sec:results}

The initial configuration for all \bbh{} systems have the holes separated by a
coordinate distance $d=8\,M_0$. The scalar field momentum shell has radius $r_0=12\,M_0$ and thickness $\sigma=1\,M_0$. We also set the mass of  the scalar field to $m_\phi = 0.4/M_0$.
Each simulation was carried out with 8 levels of mesh refinements, outer boundary at $317.44\,M_0$, and resolution in the finest grid of $M_0/64.5$.

We considered two types of binaries. One is binaries with non-spinning \bh{s} and initial mass ratios $q_0=m_1/m_2=(2,~3,~4)$. The other type is binaries with equal mass \bh{s} and their spins anti-aligned spins with magnitudes $a=0.6$. For the spinning cases, we investigated two setups: one with the \bh{} spins aligned with the orbital momentum (non-precessing binaries) and spins in the orbital plane (super-kick configuration). With the exception of the super-kick configuration binaries, we considered initial amplitude values of the scalar momentum  $\widehat \Pi_0 \equiv \Pi_0\,M_0\times 10^3=(5.0,~7.5,~10.0)$. On the other hand, for super-kick binaries, we have added more cases and set $\widehat\Pi_0=(1.25,2.5,3.75,5.0,6.25,7.5,8.50,10.0)$. In order to do comparisons with the vacuum case, we did simulations with $\widehat \Pi_0=0$ for all types. The labeling of the simulations is as follows: A non-spinning, $q_0=$ x with $\widehat\Pi_0=$ y.y simulation is labeled $q$x-0yy. Similarly, an equal mass simulation with spins perpendicular and parallel to the orbital angular momentum with the same $\widehat \Pi_0$ are labeled $a_\perp$0yy and $a_\parallel$0yy, respectively.

Tables~\ref{table:case1} and \ref{table:case2} show the scalar field energies $E_\phi$ and total ADM energy $E_{ADM}$ in the initial data for each of the cases. Notice that $E_{ADM} \simeq E_{ADM}^{vac}+ E_\phi$ where~\cite{William2017}
\begin{equation}
E_\phi=\int\rho\sqrt{\gamma}\,d^3x=\frac{1}{2}\int\widetilde\Pi^2\psi^{-6}\sqrt{\eta}\,d^3x\,.
\label{sphericalharmonic}
\end{equation}

\begin{table}
	\begin{center}
		\begin{tabular}{ c  c c}
			\hline
			\hline
			~Case~&$E_\phi/M_0$&$E_{ADM}/M_0$   \\
			\hline
			q2-000   & 0.0000 & 0.989  \\
			q2-050   & 0.0289 & 1.018  \\
			q2-075   & 0.0643 & 1.053  \\
			q2-100   & 0.1126 & 1.102  \\
			\hline
			q3-000   & 0.0000 &  0.991 \\
			q3-050   & 0.2889 &  1.019 \\
			q3-075   & 0.0643 &  1.055 \\
			q3-100   & 0.1126 &  1.104 \\
			\hline
			q4-000   & 0.0000 &  0.992 \\
			q4-050   & 0.2889 &  1.021 \\
			q4-075   & 0.0643 &  1.056 \\
			q4-100   & 0.1126 &  1.105 \\
			\hline
			\hline
		\end{tabular}
	\end{center}
				\caption{ADM and scalar field energies in the initial data for un-equal mass, non-spinning \bbh{} configurations.}
							\label{table:case1}
\end{table}

\begin{table}
	\begin{center}
		\begin{tabular}{ c  c  c  }
			\hline
			\hline
			~Case~& $E_\phi/M_0$& $E_{ADM}/M_0$   \\
			\hline
			$a_\parallel$000  & 0.0000 &  0.987          \\
			$a_\parallel$050  & 0.0289   &  1.016           \\
			$a_\parallel$075  & 0.0643  &  1.052           \\
			$a_\parallel$100  & 0.1127  &  1.101           \\
			\hline
			$a_\perp$0000     & 0.0000  &  0.987      \\
			$a_\perp$0125     & 0.0018  &  0.989      \\
			$a_\perp$0250     & 0.0072 &  0.995      \\
			$a_\perp$0375     & 0.0163  &  1.004      \\
			$a_\perp$0500     & 0.0289 &  1.017      \\
			$a_\perp$0625     & 0.0449 &  1.033      \\
			$a_\perp$0750     & 0.0643  &  1.052      \\
			$a_\perp$0875     & 0.0869 &  1.075     \\
			$a_\perp$1000     & 0.1127 &  1.101      \\
			\hline
			\hline
		\end{tabular}
	\end{center}
				\caption{ADM and scalar field energies in the initial data for equal mass, spinning \bbh{} configurations.}
							\label{table:case2}
\end{table}

\section{Un-equal Mass, Non-spinning \bh{} Binaries}
\label{sec:results1}

\begin{figure*}
	\includegraphics[width=0.32\linewidth]{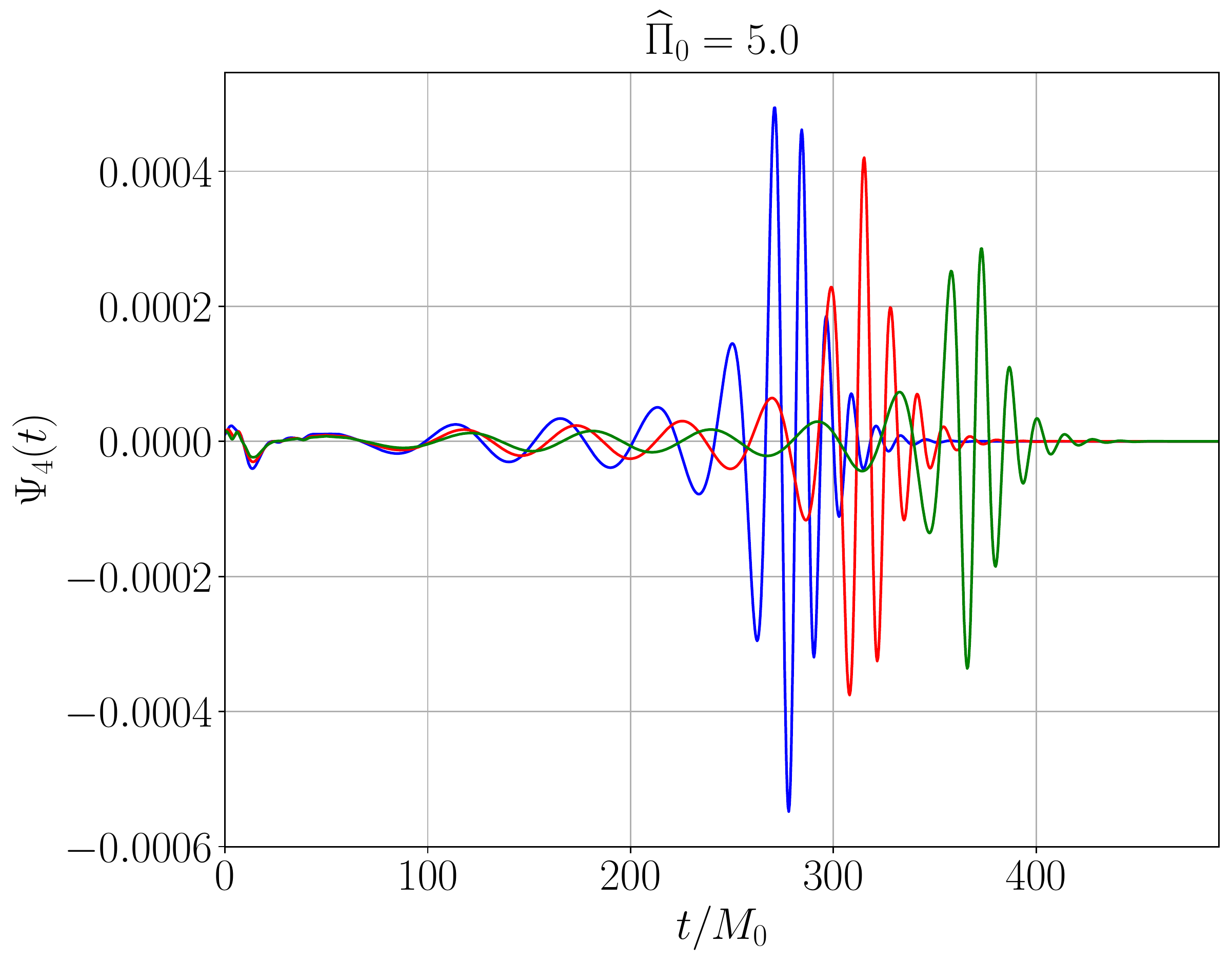}
	\includegraphics[width=0.32\linewidth]{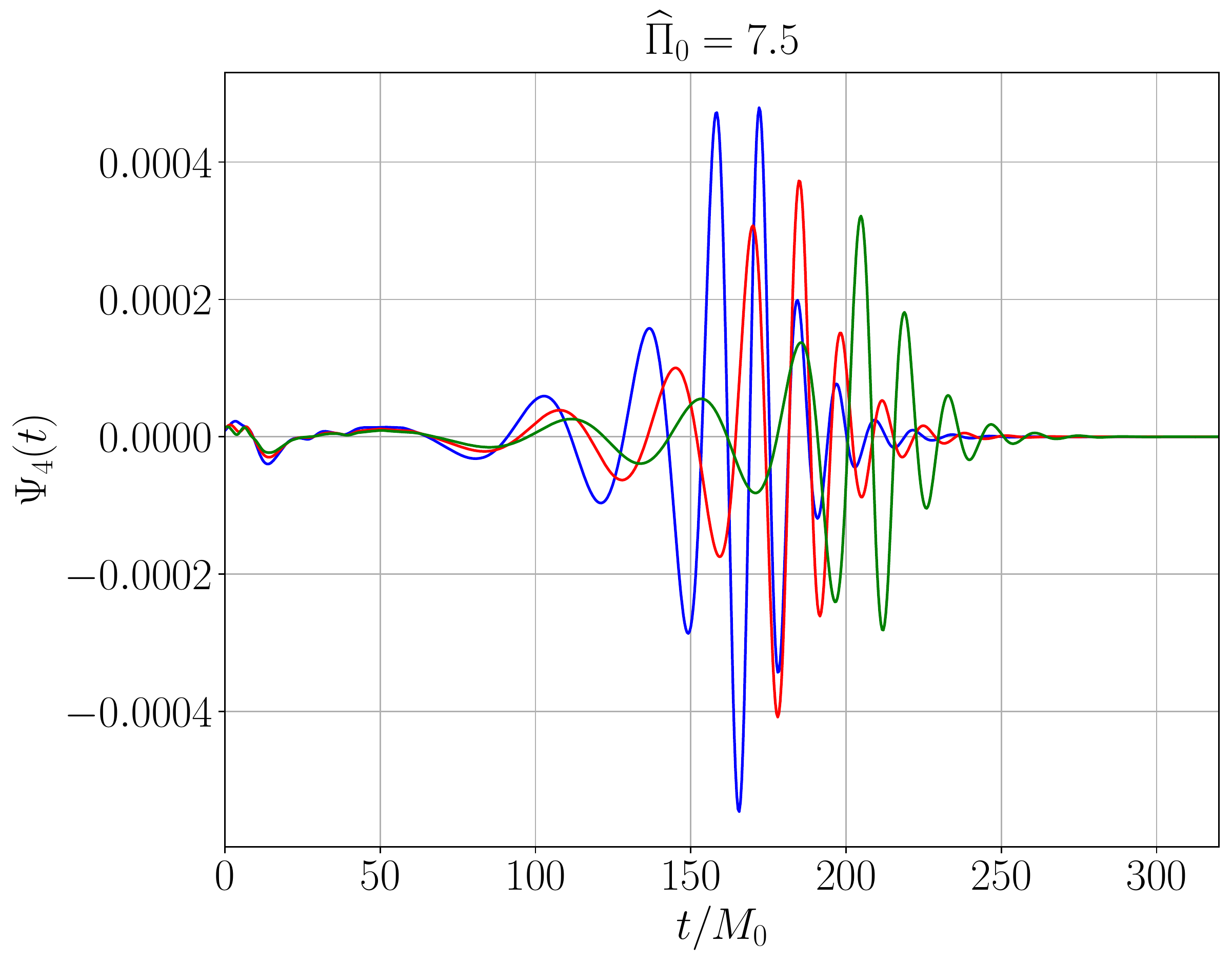}
	\includegraphics[width=0.32\linewidth]{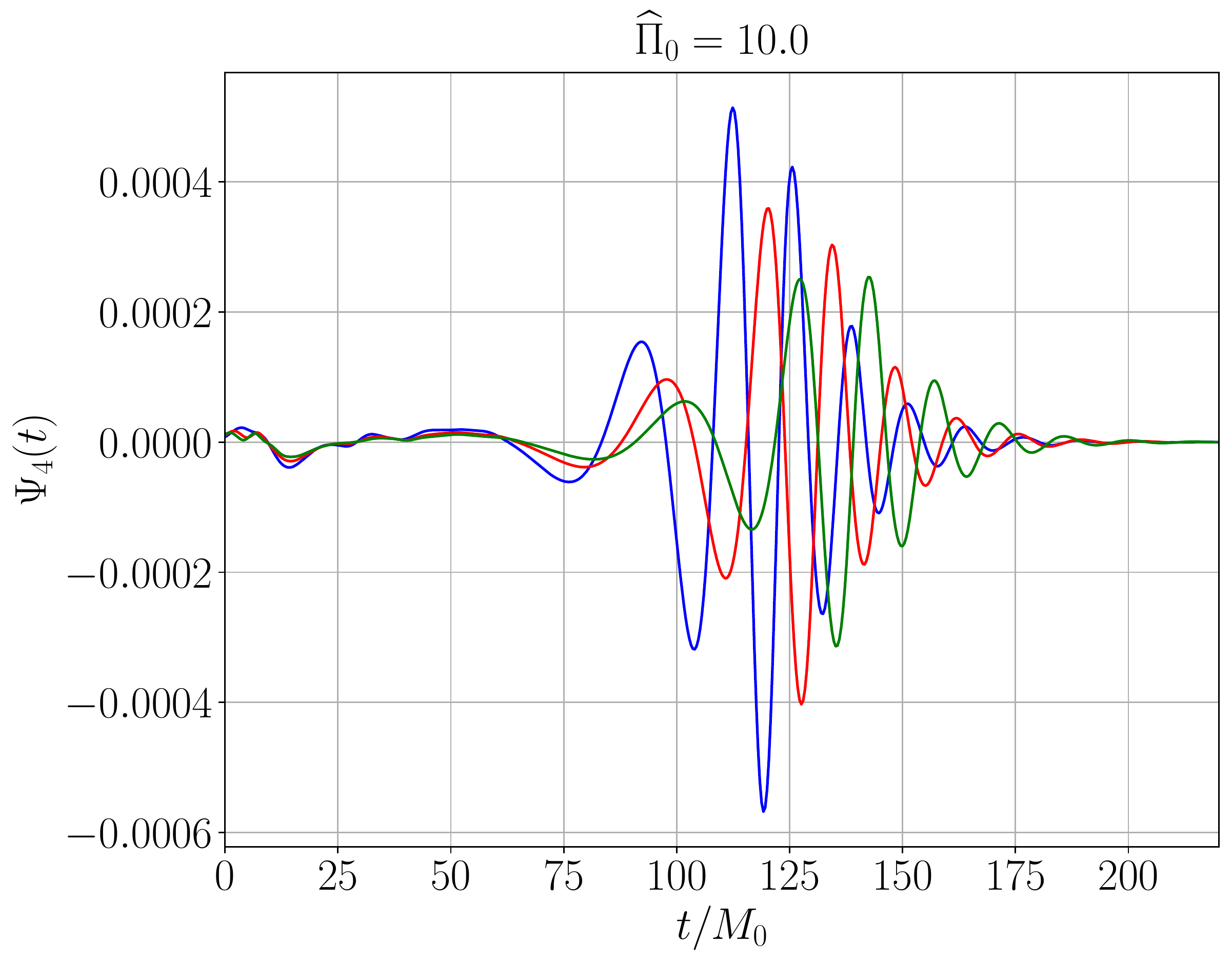}
	\includegraphics[width=0.32\linewidth]{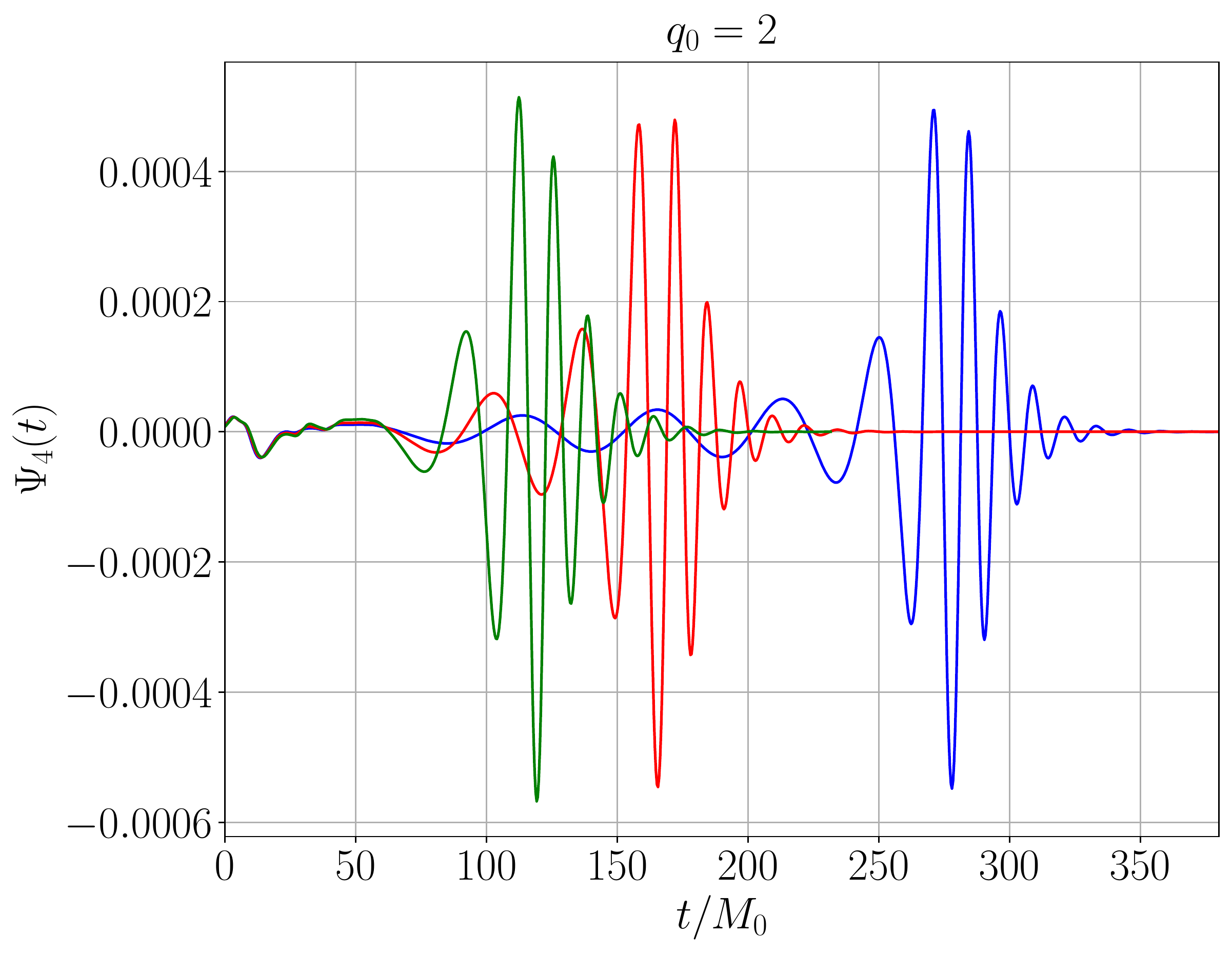}
	\includegraphics[width=0.32\linewidth]{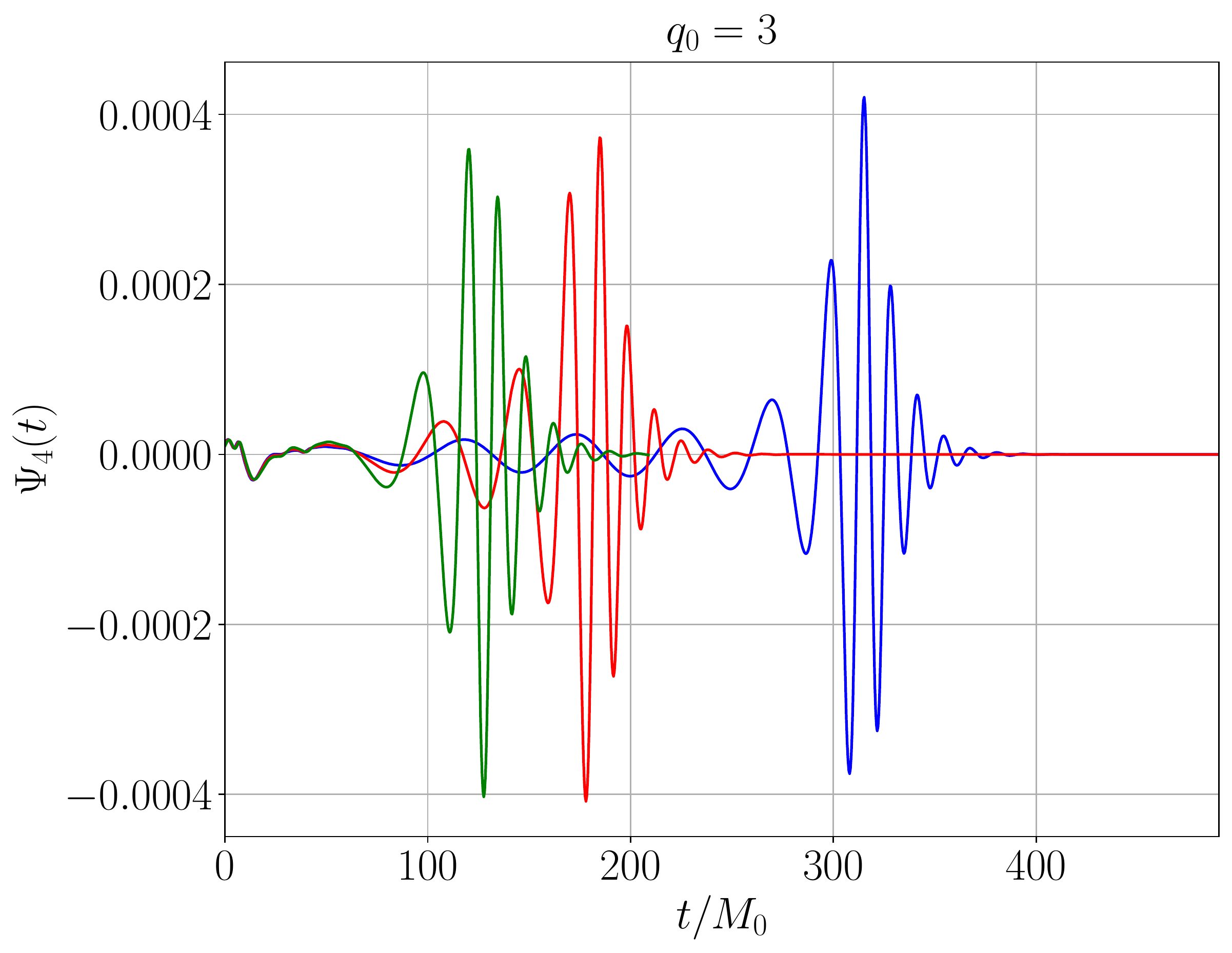}
	\includegraphics[width=0.32\linewidth]{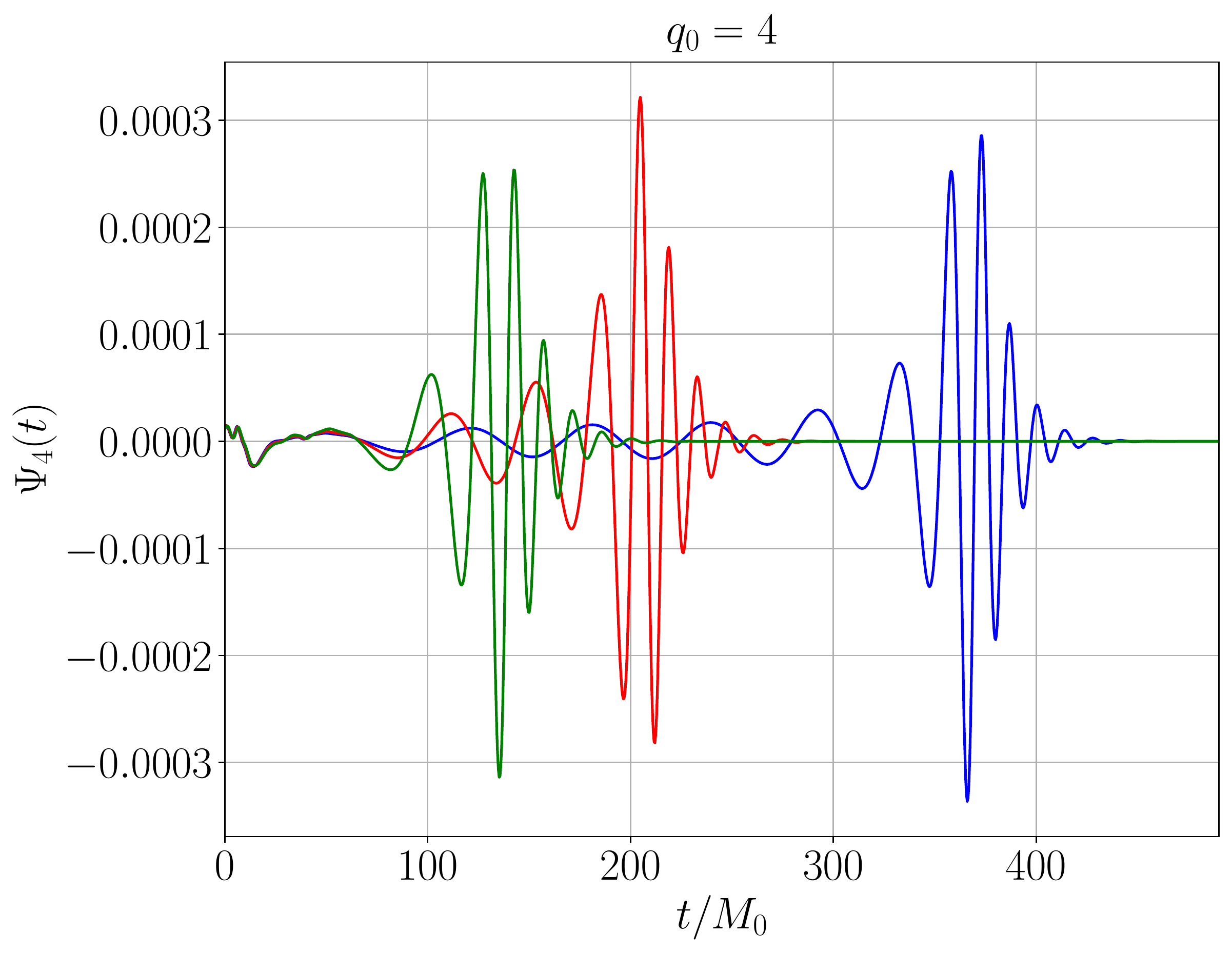}
	\caption{Mode $l=2$, $m=2$ of the Weyl scalar $\Psi_4$ for the un-equal mass and non-spinning \bh{} binaries.  The top panels from left to right are for  $\widehat\Pi_0= (5.0,~7.5,~10.0)$, respectively, with lines blue, red, and green corresponding to $q_0=(2,~3,~4)$, respectively. The bottom panels from left to right are for  $q_0=(2,~3,~4)$, respectively, with lines blue, red, and green corresponding to $\widehat\Pi_0= (5.0,~7.5,~10.0)$, respectively. }
	\label{fig:gw1}
\end{figure*}

Figure~\ref{fig:gw1} shows the mode $l=2$, $m=2$ of the Weyl scalar $\Psi_4$ for the un-equal mass and non-spinning \bh{} binaries. The top panels from left to right are for  $\widehat\Pi_0= (5.0,~7.5,~10.0)$, respectively, with lines blue, red, and green corresponding to $q_0=(2,~3,~4)$, respectively. The bottom panels from left to right are for  $q_0=(2,~3,~4)$, respectively, with lines blue, red, and green corresponding to $\widehat\Pi_0= (5.0,~7.5,~10.0)$, respectively.
From the top panels we see that, for a given $\widehat\Pi_0$, the binary merges earlier for smaller $q_0$, as expected from the vacuum case, since the luminosity in \gw{} during the inspiral scales as $q^2/(1+q)^4$~\cite{baumgarte_shapiro_2010}. At the same time, for a given $q_0$, the larger the given value of $\widehat\Pi_0$ is, the smaller the difference among the merger times.

From the bottom panels in Fig.~\ref{fig:gw1}, one sees that for a given $q_0$, the larger $\widehat\Pi_0$, the earlier the binary merges. This is because the luminosity in \gw{} also depends on the total mass of the binary $M$ as $M^2$~\cite{baumgarte_shapiro_2010}. And as we shall see next, $M$ grows monotonically with $\widehat\Pi_0$. Also, when one slices the data this way, we observe that the differences with $\widehat\Pi_0$ in merger times remain roughly the same independently of $q_0$.

\begin{figure*}
	\includegraphics[width=0.49\linewidth]{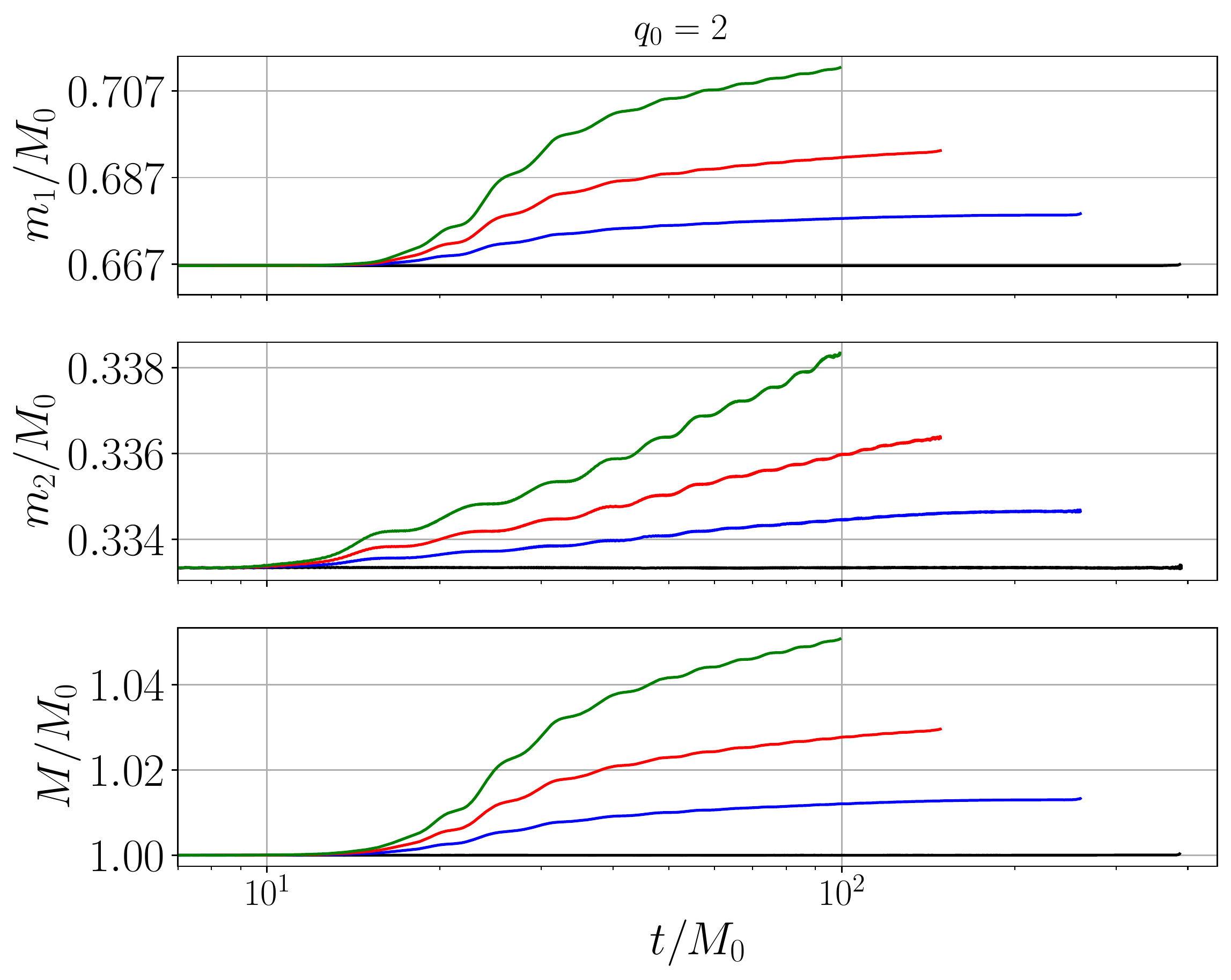}
	\includegraphics[width=0.49\linewidth]{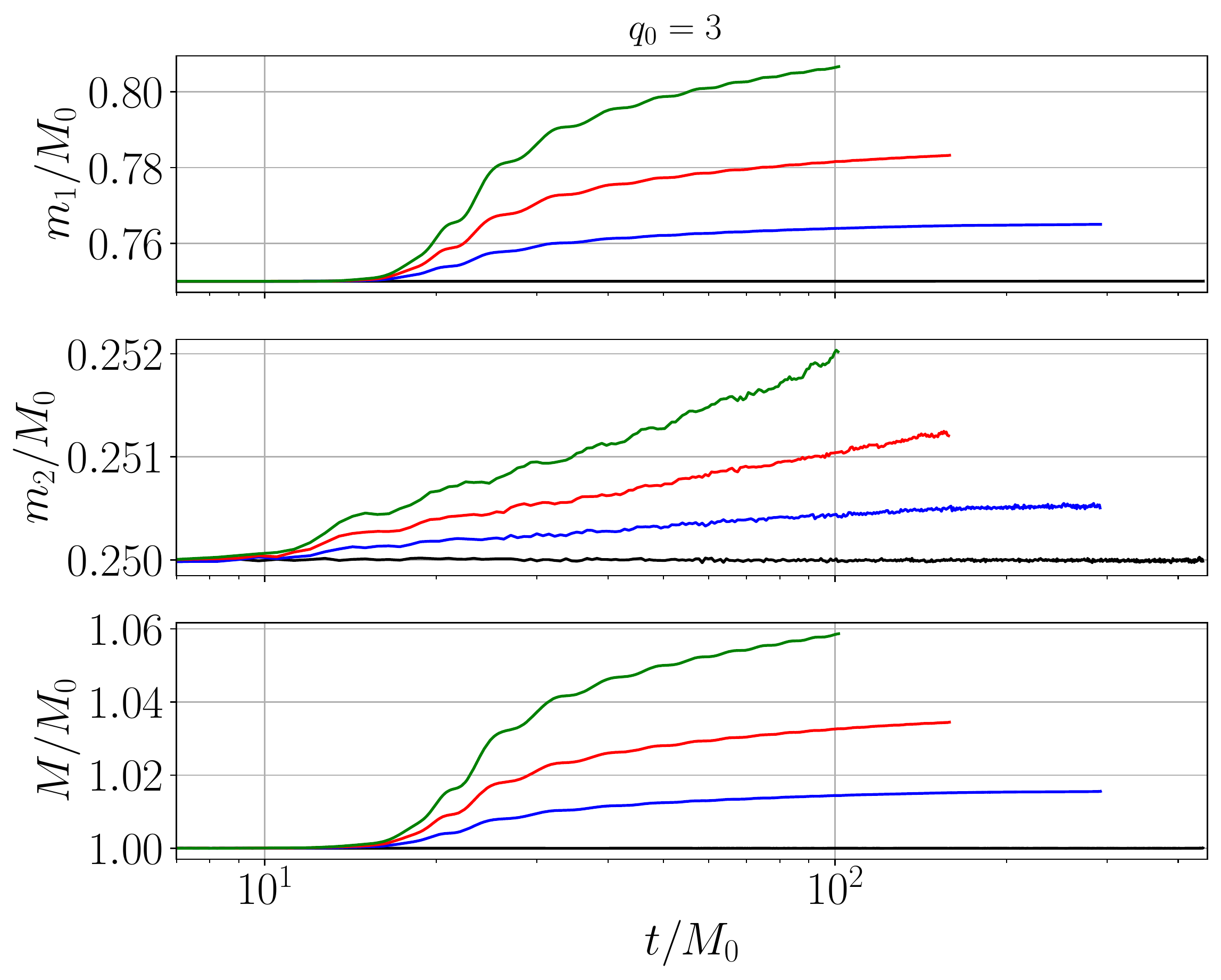}
	\includegraphics[width=0.49\linewidth]{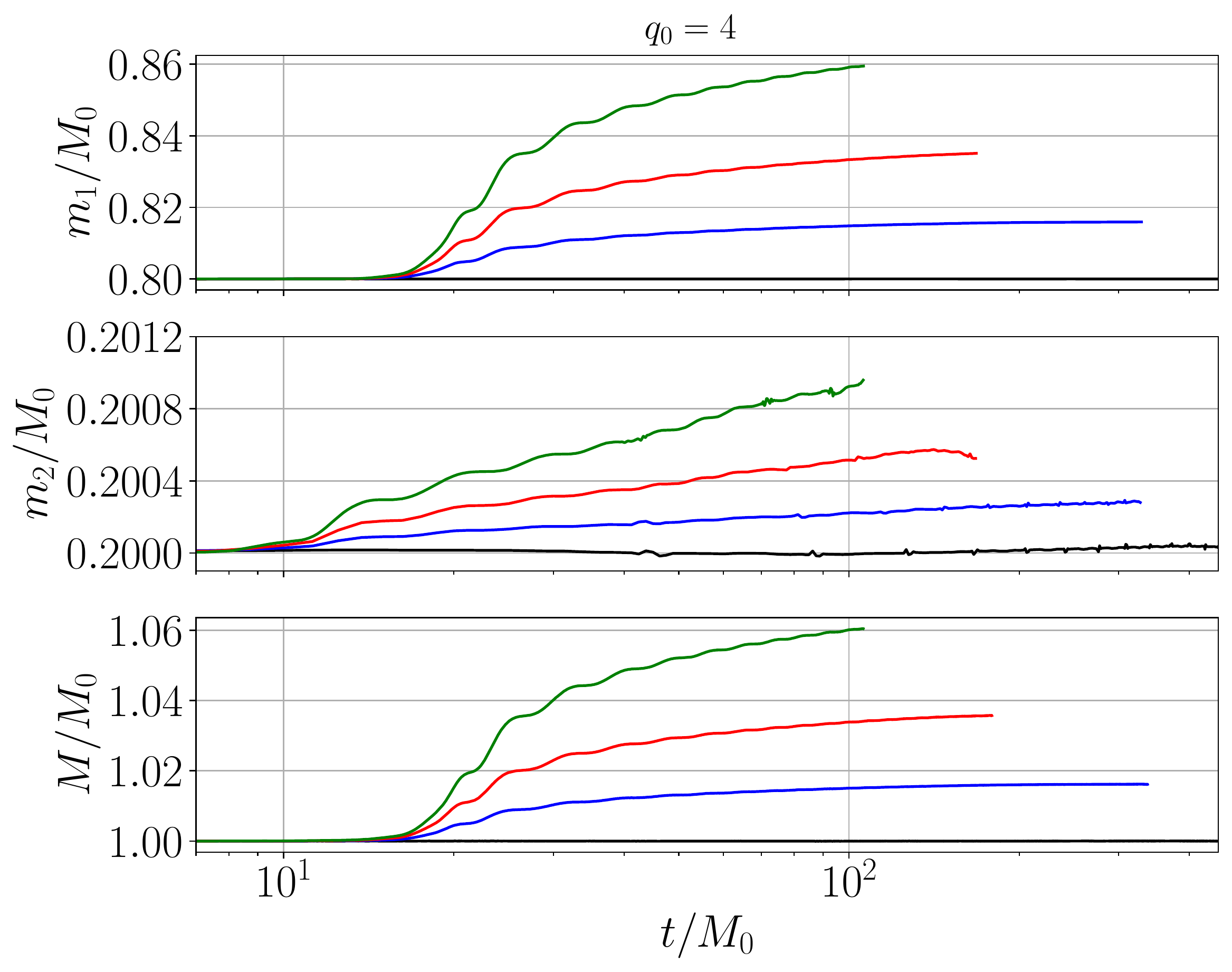}
	\includegraphics[width=0.465\linewidth]{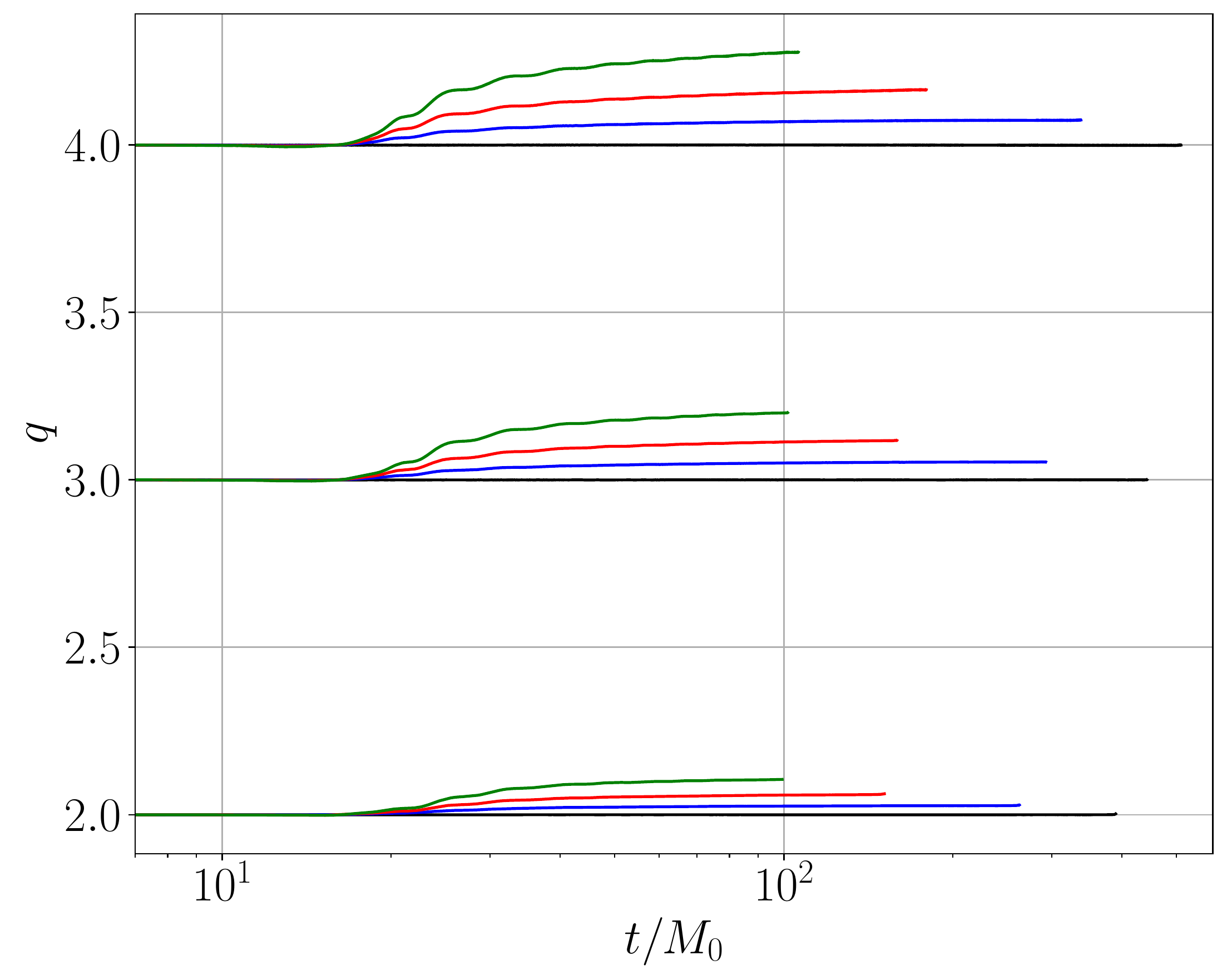}
	\caption{Un-equal mass, non-spinning \bbh{s}: Evolution of the \bh{} masses $m_1$ and $m_2$, the total mass $M$, and the mass ratio $q$ due to scalar field accretion. The black, red, blue, and green correspond to $\widehat\Pi_0=(0,~5.0,~7.5,~10.0)$, respectively. The lines end at the time when the merger occurs.}
	\label{fig:masses1}
\end{figure*}

\begin{figure*}
	\includegraphics[width=0.41\linewidth]{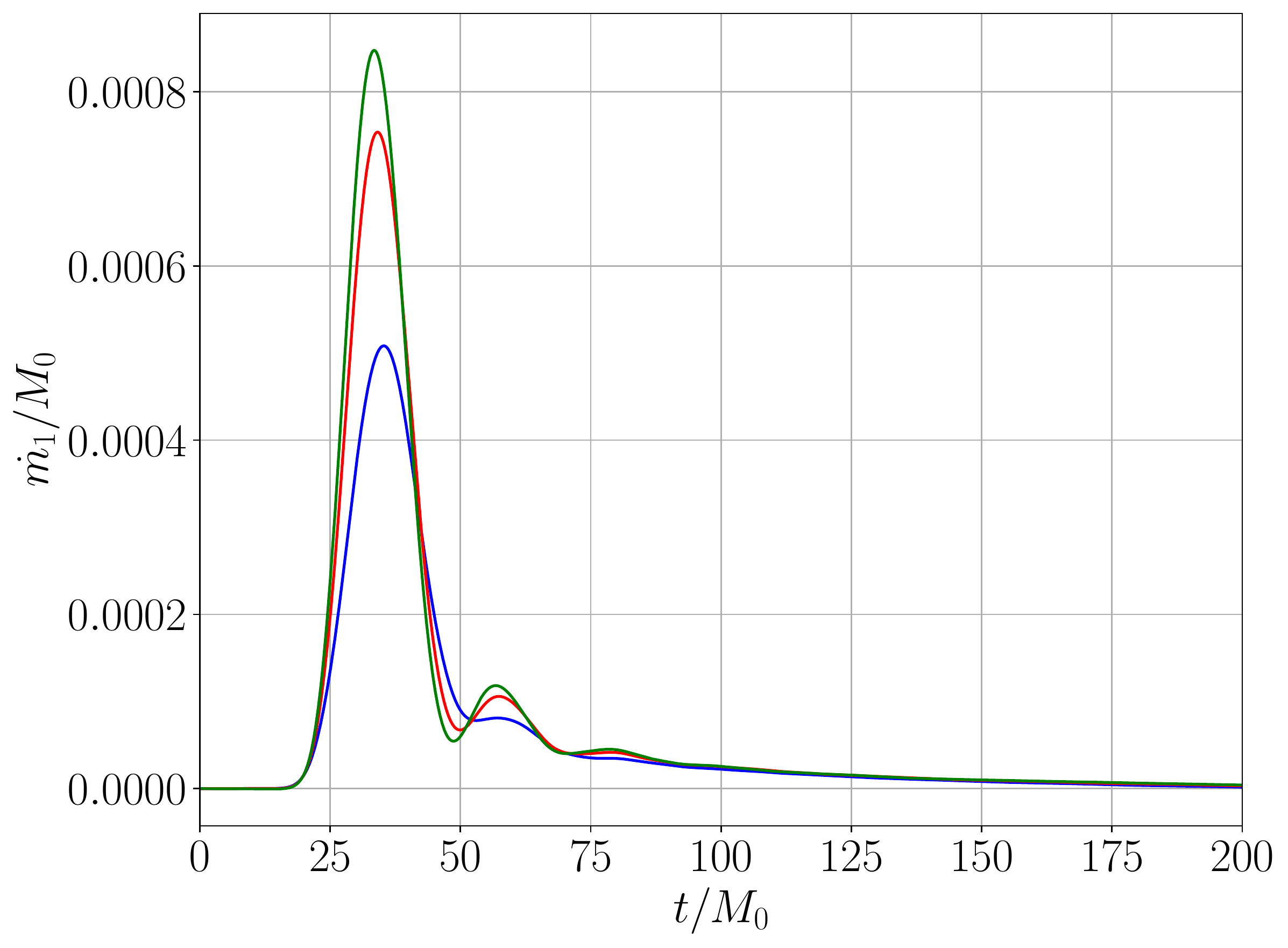}
	\includegraphics[width=0.4\linewidth]{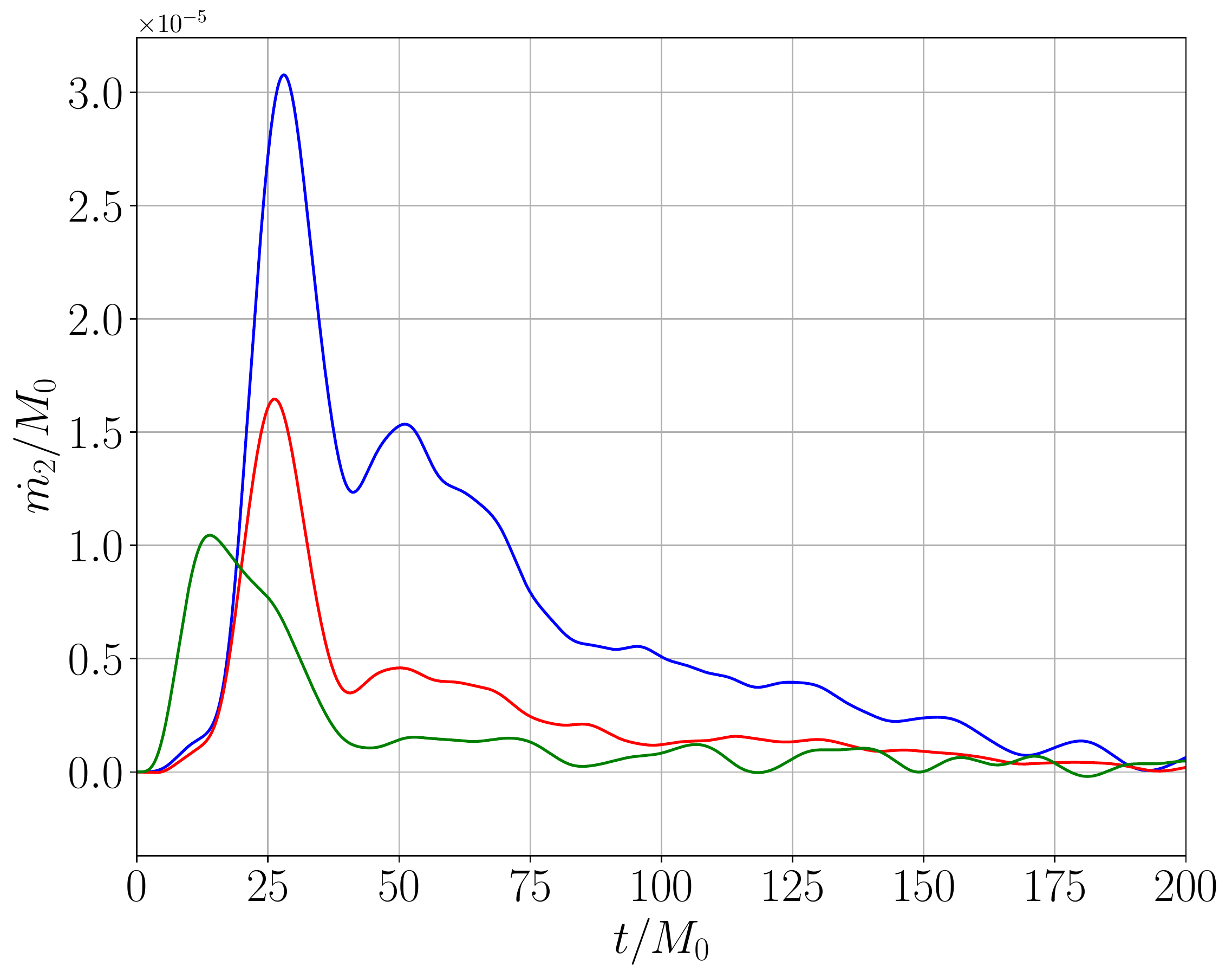}
	\includegraphics[width=0.41\linewidth]{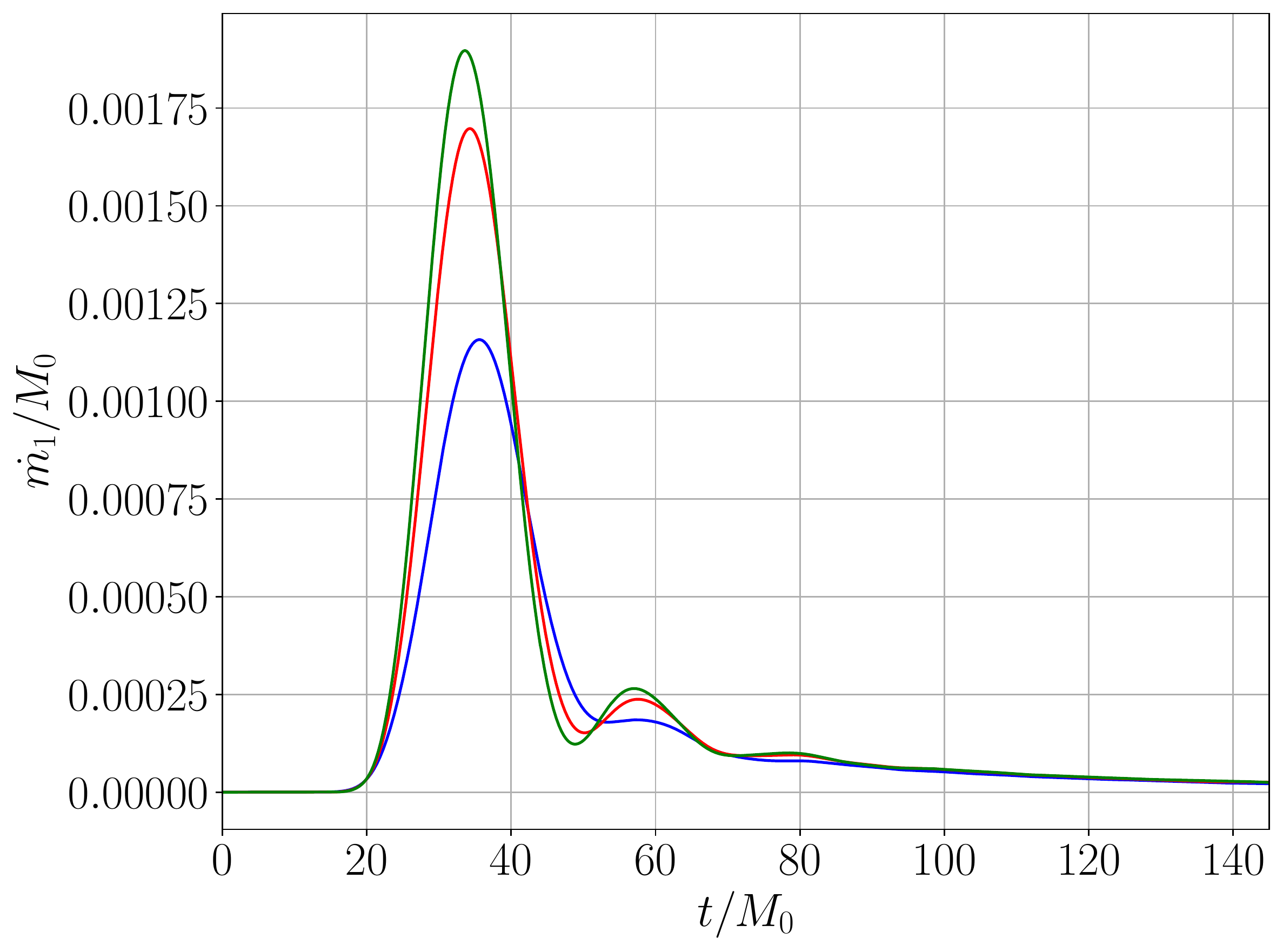}
	\includegraphics[width=0.38\linewidth]{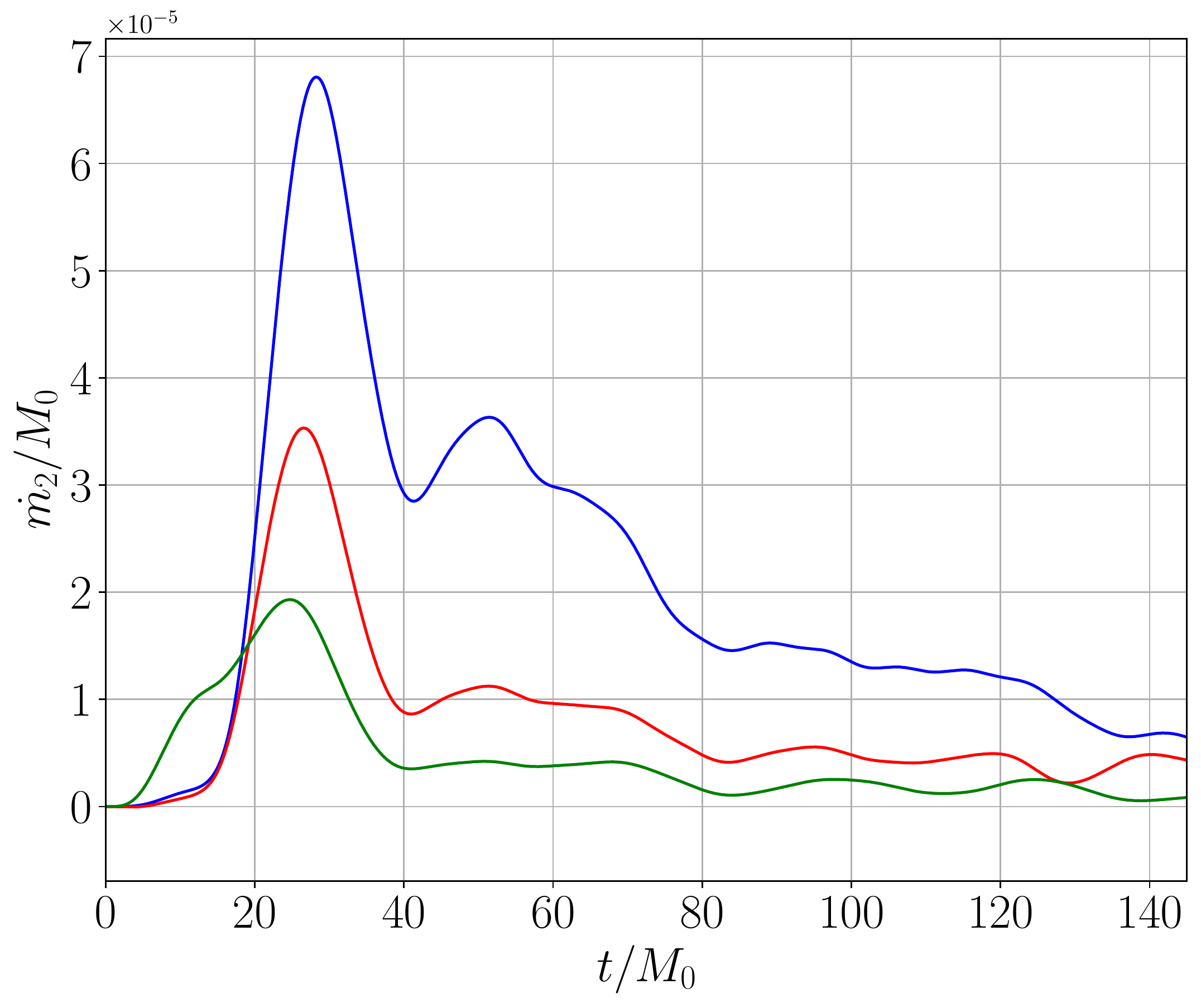}
	\includegraphics[width=0.41\linewidth]{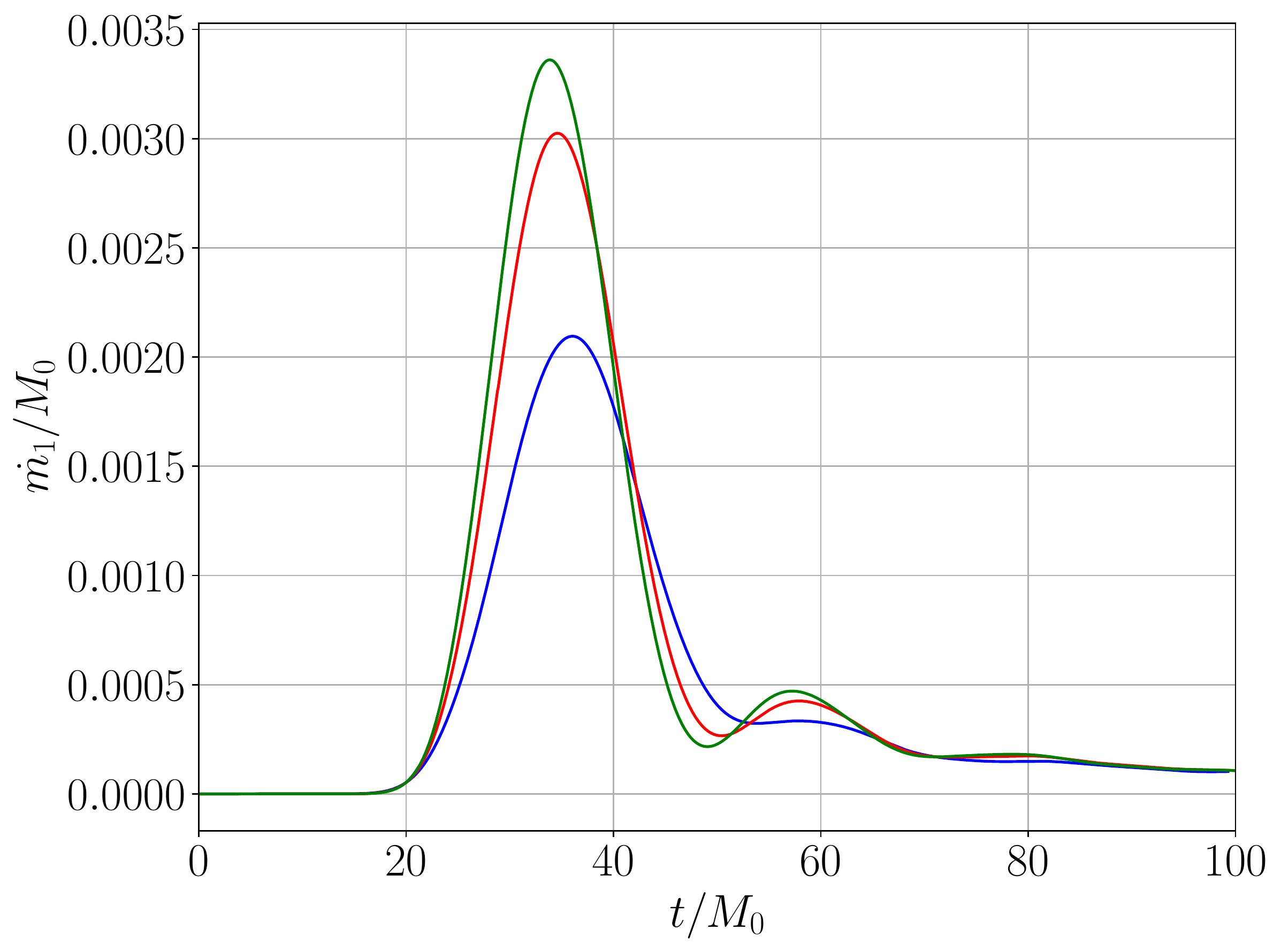}
	\includegraphics[width=0.41\linewidth]{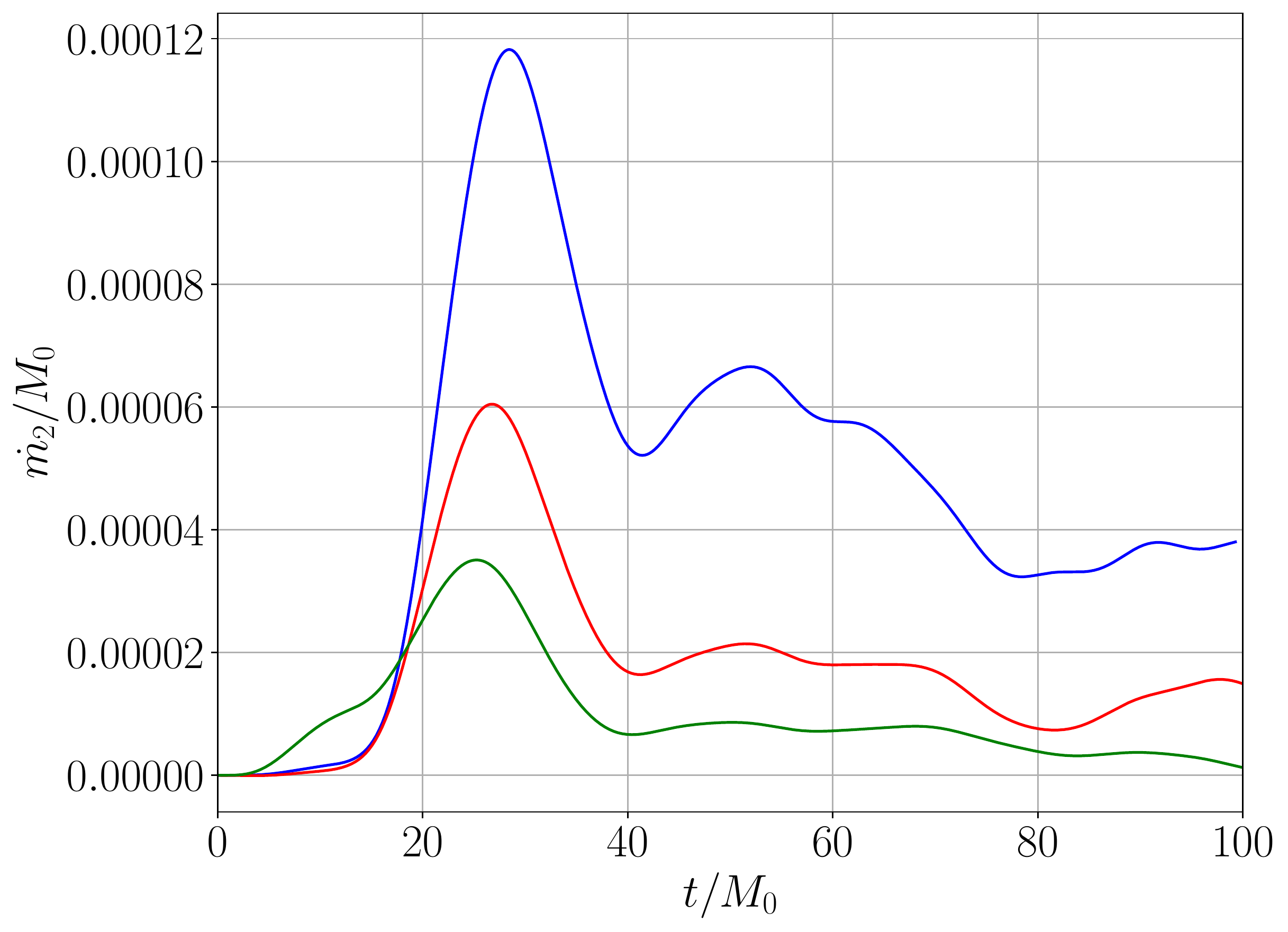}
	\caption{Un-equal mass, non-spinning \bbh{}: Mass accretion rates for each \bh{}, top to bottom panels $\widehat\Pi_0=(0,~5.0,~7.5,~10.0)$, blue, red, and green correspond to $q_0=(2,~3,~4)$, respectively.}
	\label{fig:dmasses1dt}
\end{figure*}

The accretion of the scalar field by the \bh{s} modifies the total binary mass $M$ and its mass ratio $q$ as it evolves.  Figure~\ref{fig:masses1} shows the evolution of $m_1$, $m_2$, and $M$ for each initial $q_0$. As expected, accretion starts when the scalar field shell reaches the \bh{s}, approximately at a time $\sim r_0$. The bottom right panel also shows the evolution of $q$ due to the changes of the \bh{} masses. In all panels, lines terminate at the time when the binary mergers, as signaled by the appearance of a common apparent horizon. The colors black, blue, red and green denote $\widehat\Pi_0=(0,~5.0,~7.5,~10.0)$, respectively. Figure~\ref{fig:dmasses1dt} shows the corresponding \bh{} accretion rates.

From Figs.~\ref{fig:masses1} and~\ref{fig:dmasses1dt}, we observe that the \bh{} masses and accretion rates grow monotonically with $\widehat \Pi_0$ for a given initial $q_0$. Furthermore, the growth is such that the increase in $q$ is also monotonic with $\widehat \Pi_0$. From Fig.~\ref{fig:dmasses1dt}, given a value of $\widehat \Pi_0$, $\dot m_1 > \dot m_2$, similar to Bondi accretion behaviour in which the accretion rate is proportional to the mass of the accreting object. By taking into consideration the growth in $q$ observed in Fig.~\ref{fig:masses1}, namely $\dot q > 0$, one obtains that $\dot m_1 > q\,\dot m_2$. 

\begin{figure*}
	\includegraphics[width=0.32\linewidth]{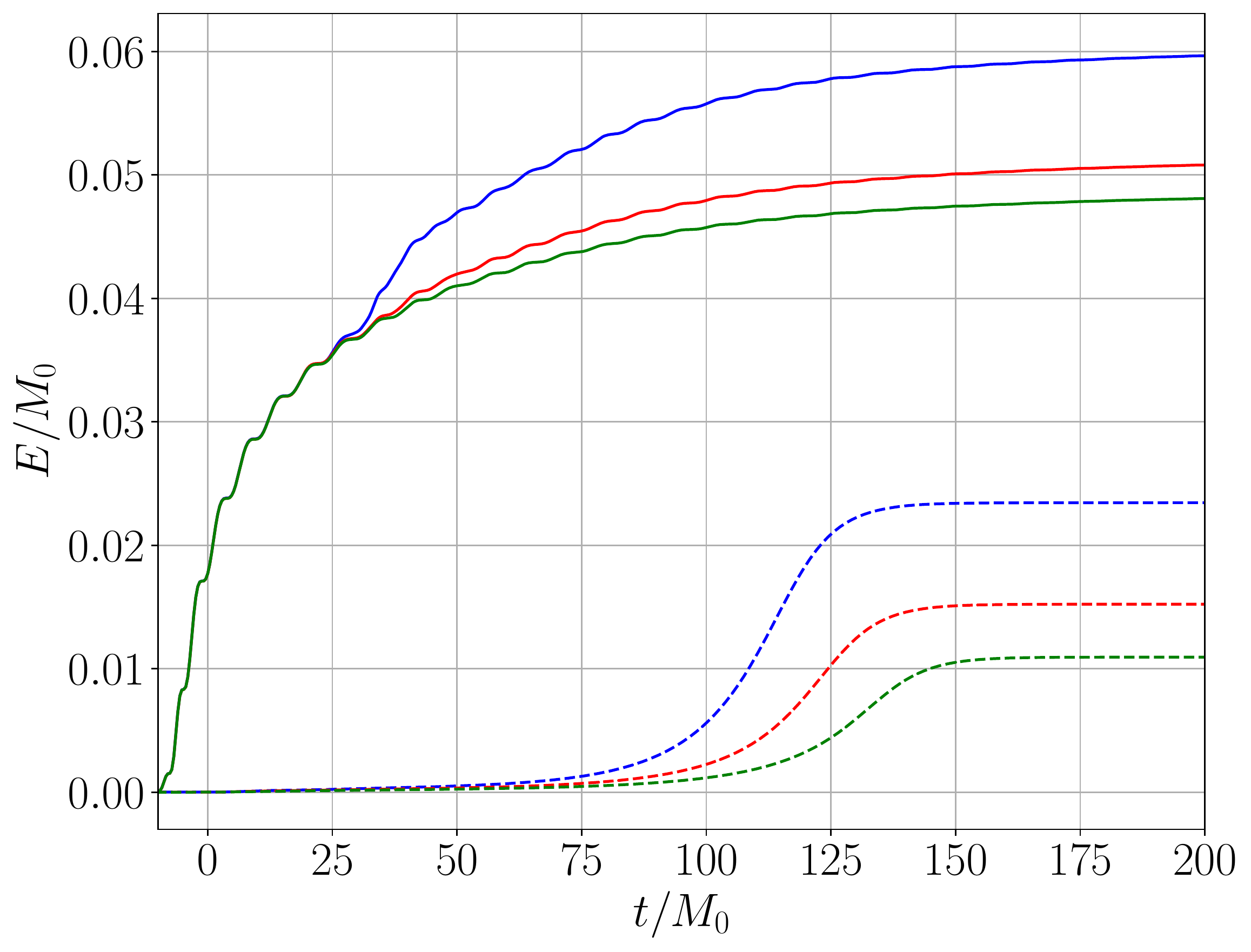}
	\includegraphics[width=0.32\linewidth]{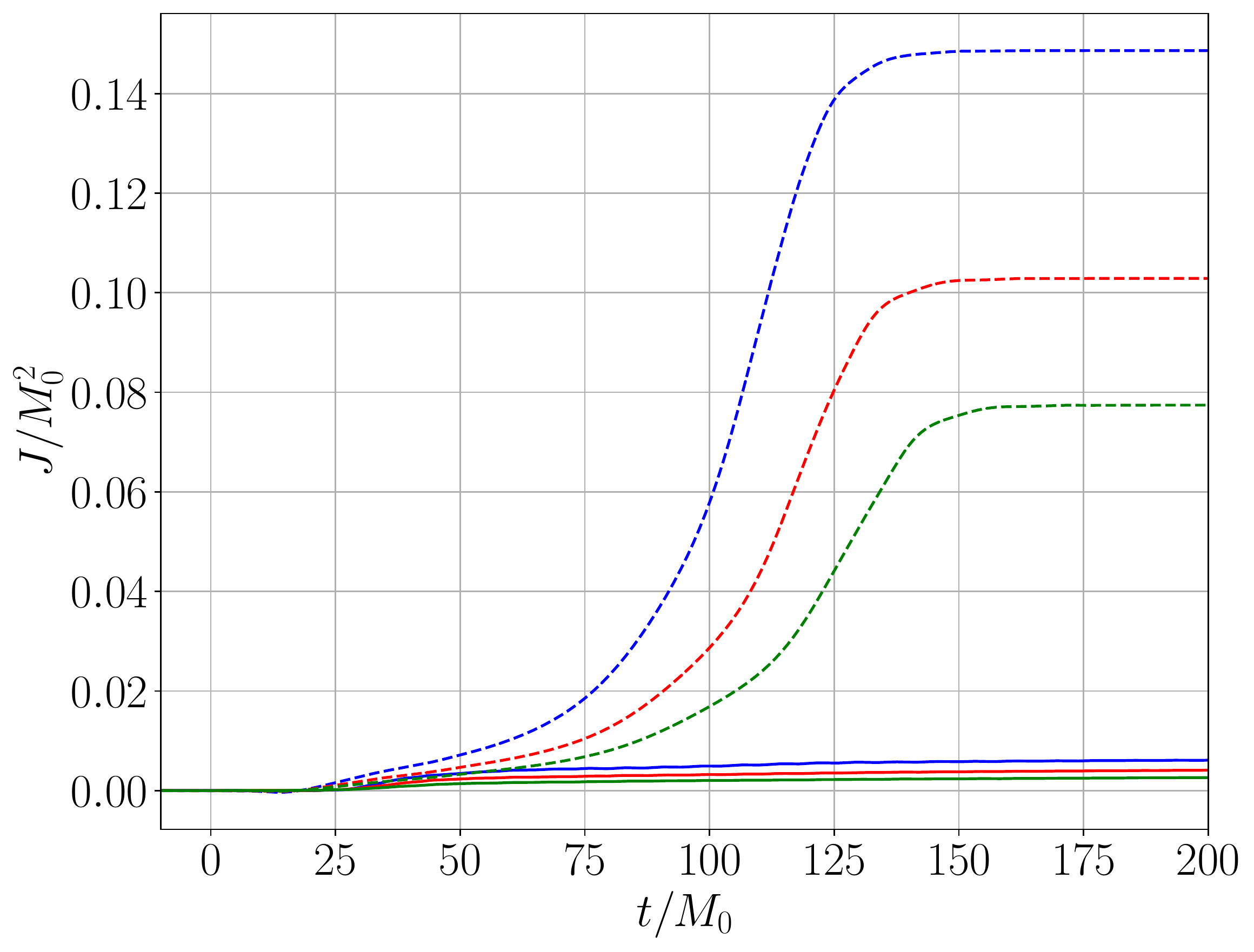}
	\includegraphics[width=0.324\linewidth]{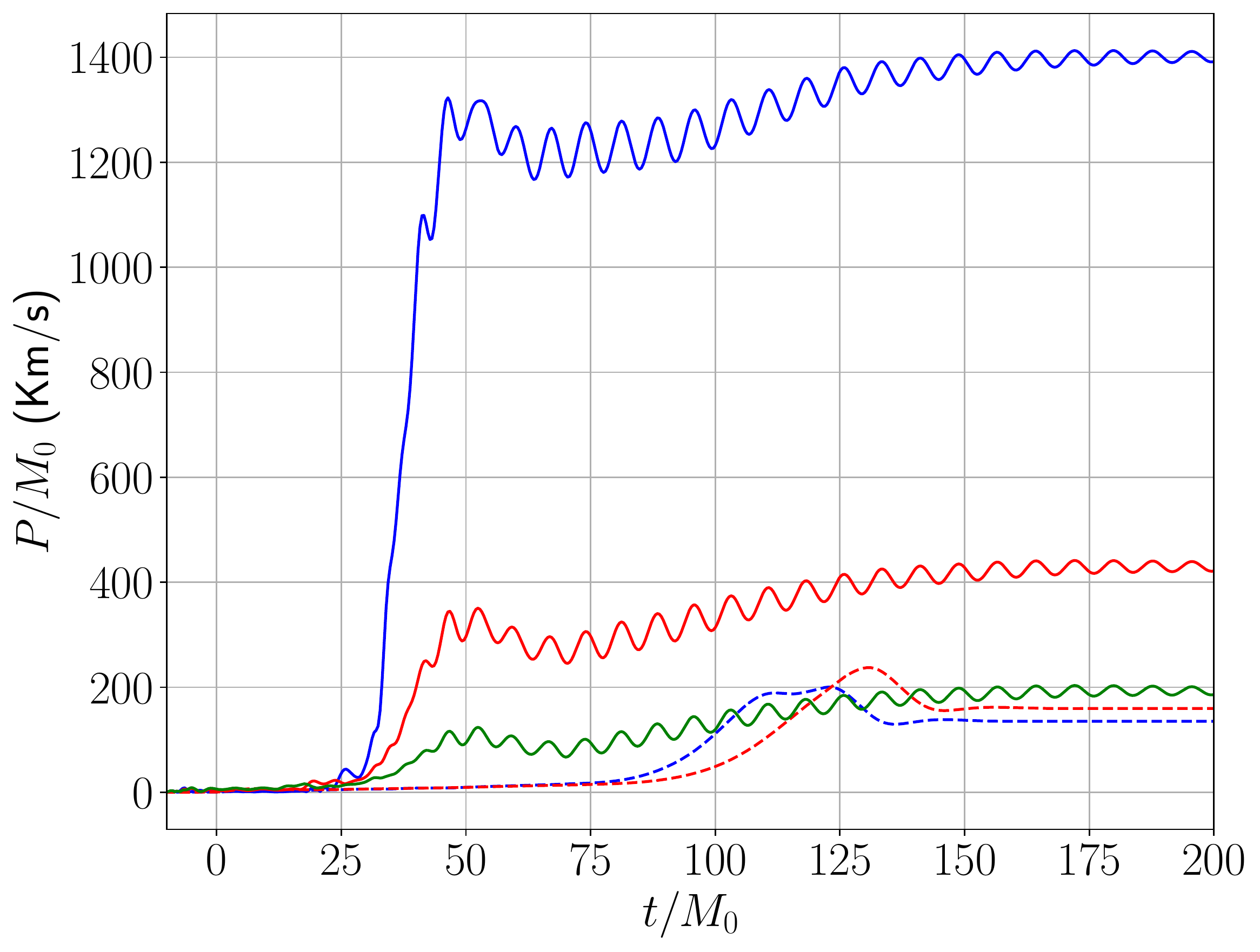}
	\caption{Un-equal mass, non-spinning \bbh{}: energy, angular momentum, and linear momentum radiated in \gw{s} (dashed lines) and in the scalar field (solid lines) for the case $\widehat\Pi_0=10.0$. Colors blue, red and green correspond to $q_0=(2,~3,~4)$, respectively.}
	\label{fig:radn1}
\end{figure*}

 Figure~\ref{fig:radn1} shows the energy, angular momentum, and linear momentum radiated in \gw{s} (dashed lines) and in the scalar field (solid lines) for the case $\widehat\Pi_0=10.0$. We observe in the left panel that the energy radiated by the scalar field is higher than in \gw{s}. This can be explained as follows: the ADM energy at the end of the simulations is given by $E_{ADM} = E^{rad}_{\gw{}}+E^{rad}_\phi + m_f$, with $m_f$ the mass of the final \bh{}. For the case $q_0=2$ and $\widehat\Pi_0 = 10.0$, we have from Table~\ref{table:case1} that $E_{ADM} =1.102\,M_0 $ and from Table~\ref{table:kick-velocity-table} that $m_f = 1.0147\, M_0$; thus,
 $E^{rad}_{\gw{}}+E^{rad}_\phi = E_{ADM} -m_f \simeq 0.087\,M_0$. Since energy radiated in \gw{s} is typically a few percent, in this case $E_{\gw{}} \simeq 0.025\,M_0$, we have that $E^{rad}_\phi \simeq 0.06\,M_0$, consistent with the value in  Fig.~\ref{fig:radn1}. Another characteristic in this figure is that, as with \gw{s}, the energy radiated in the scalar field decreases monotonically with $q_0$.

 The angular momentum radiated is depicted in the middle panel of Fig.~\ref{fig:radn1}. As expected, \gw{s} carry away angular momentum and shrink the binary. The scalar field also extracts angular momentum but in smaller amounts. The reason why the scalar field angular momentum radiation is much smaller than the one in \gw{s} is because initially the scalar field shell does not have any angular momentum. All the momentum generated is from the ``stirring" of the scalar field by the binary.
 
The right panel in Fig.~\ref{fig:radn1} shows the magnitude of linear momentum emitted, which for these non-precessing binaries lies in the $xy$-plane. As with the energy radiated, the emission of scalar field linear momentum is significantly larger than in the \gw{s}. Also interesting is the oscillations in the scalar field linear momentum radiated, which are also observed in the energy and angular momentum but at a much smaller scale. The reason for this is because in systems of \bbh{} with massive scalar fields, as it is in our case, the scalar fields develops long-lived modes due to the presence of an effective potential. 

Tables~\ref{table:kick-velocity-table} shows the mass $m_f$, spin $a_f$, and kick velocity $v_{kick}$ of the remnant \bh{,} where we have combined the emission of linear momentum by \gw{s} and the scalar field to estimate the gravitational recoil. Independently of $q_0$, $m_f$ grows monotonically with $\widehat\Pi_0$. This is expected from the way the \bh{s} accrete the scalar field, namely, the more massive the hole, the more it accretes. 

Regarding the final spin, we found that for a given $\widehat\Pi_0$, $a_f$ decreases as $q_0$ increases. Which is the same trend observed in the vacuum case; that is, the scalar field modifies the spin magnitude but not its dependence with $q$. On the other hand, if one fixes the attention to the final spin for a given $q_0$, one sees monotonicity in the $q_0 =3$ and 4, decreasing its value with $\widehat\Pi_0$ increasing. At first look, this seems counter intuitive because one would think that, since the larger the value of $\widehat\Pi_0$, the earlier the binary merger, there would be a larger residual of angular momentum that goes into the final spin. Yes, there is more angular momentum in the final \bh{}, but one has to also remember that $a_f = S_f/m_f^2$ is the dimensionless spin parameter, not the angular momentum $S_f$. It is the growth in the final mass of the \bh{} responsible for the decrease in $a_f$. Since the growth in the masses  for $q_0=2$ is not as large (see Fig.~\ref{fig:masses1}), the monotonicity of $a_f$ with $\widehat\Pi_0$ only shows for large values. 

For the kick velocity, given a value of $q_0$, the recoil is larger than in the vacuum case and increases monotonically with $\widehat \Pi_0$ . In vacuum, the maximum kick velocity of the final \bh{} in non-spinning, unequal-mass \bbh{} occurs near $q_0=3$~\cite{Sperhake2007}. In the presence of scalar field, we observe that the maximum kick for a given $\widehat\Pi_0$ occurs for $q_0 \le 2$, with $\widehat\Pi_0=10.0$ reaching super-kick levels. For a given $\widehat\Pi_0$, all the kicks are larger than in the vacuum case, the reason for this is because in these configurations the emission of linear momentum is larger through the scalar field channel. The initial momentum in the scalar field is not directly responsible for this since it does not have net linear momentum; it is spherically symmetric. It is through the interactions with the binary that linear momentum in the scalar field is redistributed and emitted in a particular direction. It turns out that this direction is aligned with that of the linear momentum emitted in \gw{s}.

\begin{table}[!htb]
	\begin{center}
		\begin{tabular}{c  c c c }
			\hline
			\hline
			~Case~&$m_f/M_0$&$a_f$& $v_{kick}$ (km/s)  \\
			\hline
			q2-000           & 0.9612   &  0.6232 & 146       \\
			q2-050           & 0.9743   &  0.6218   & 550        \\
			q2-075           & ~0.9893~   &  ~0.6230~   & 946         \\
			q2-100           & 1.0147   &  0.6267   & 1303       \\
			\hline
			q3-000             & 0.9712  &  0.5405    &   166   \\
			q3-050             & 0.9869  &  0.5378    & 289       \\
			q3-075             & ~1.0055~  &  ~0.5370~    & 409  \\
			q3-100             & 1.0337  &  0.5355    & 543       \\
			\hline
			q4-000             & 0.9777  &  0.4713  &  149    \\
			q4-050             & 0.9942  &  0.4686    & 202       \\
			q4-075             & ~1.0137~  &  ~0.4646~    & 256       \\
			q4-100             & 1.0422  &  0.4624    & 304       \\
			\hline
			\hline
		\end{tabular}
	\end{center}
				\caption{Mass $m_f$, spin $a_f$ and kick of the final \bh{} for the unequal mass, non-spinning \bbh{}  .}
							\label{table:kick-velocity-table}
\end{table}

\section{Equal Mass, Spinning \bh{} Binaries}
\label{sec:results2}

As mentioned before, we considered two setups for binaries with equal mass and anti-aligned spinning \bh{s}. The $a_\parallel$ cases have \bh{} spins along the direction of the orbital angular momentum (i.e., non-precessing binaries), and the $a_\perp$ cases have \bh{} spins in the orbital plane in the super-kick configuration~\cite{PhysRevLett.98.231101,PhysRevLett.98.231102}.  

\begin{figure*}[!htb]
	\includegraphics[width=0.325\linewidth]{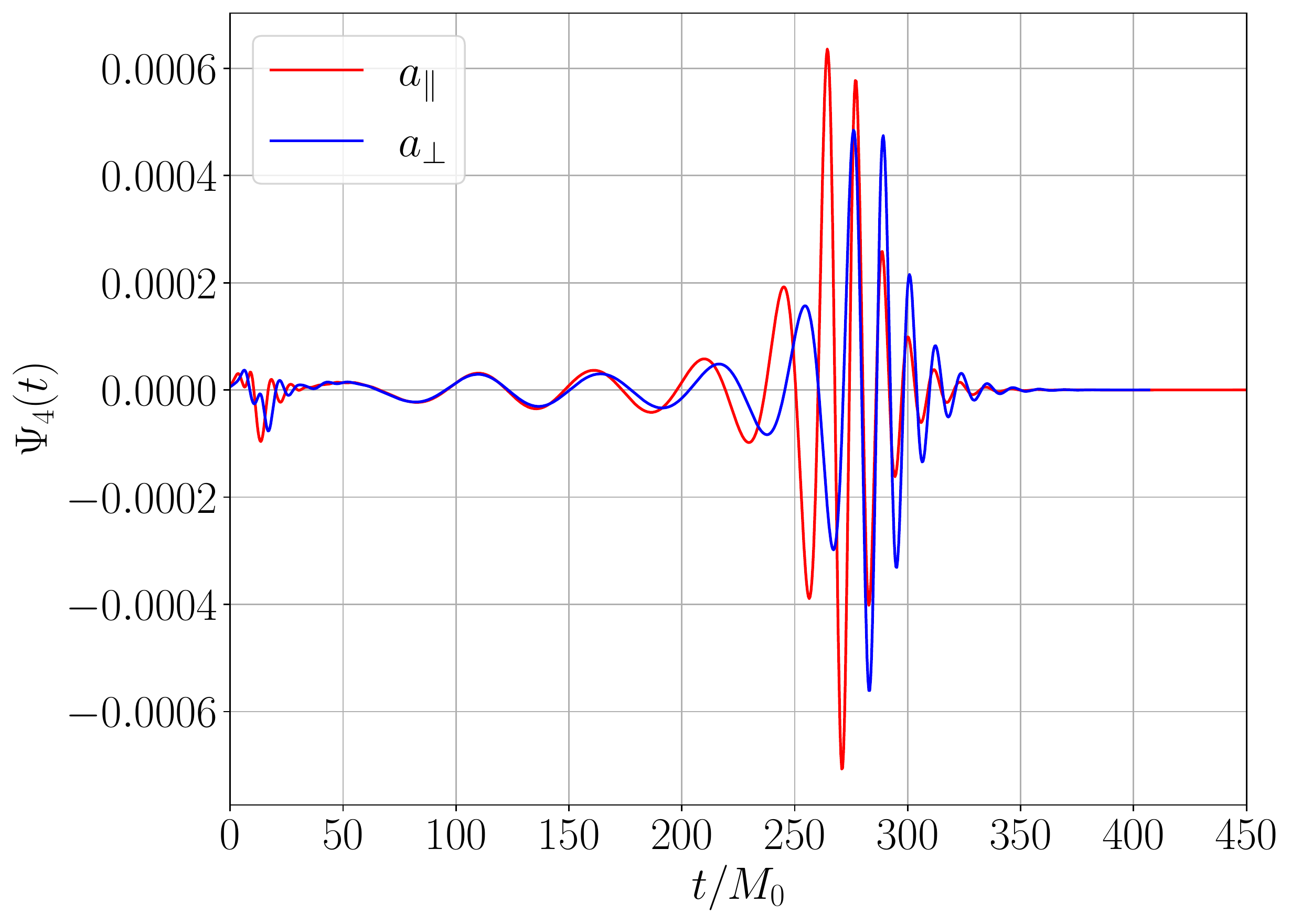}
	\includegraphics[width=0.325\linewidth]{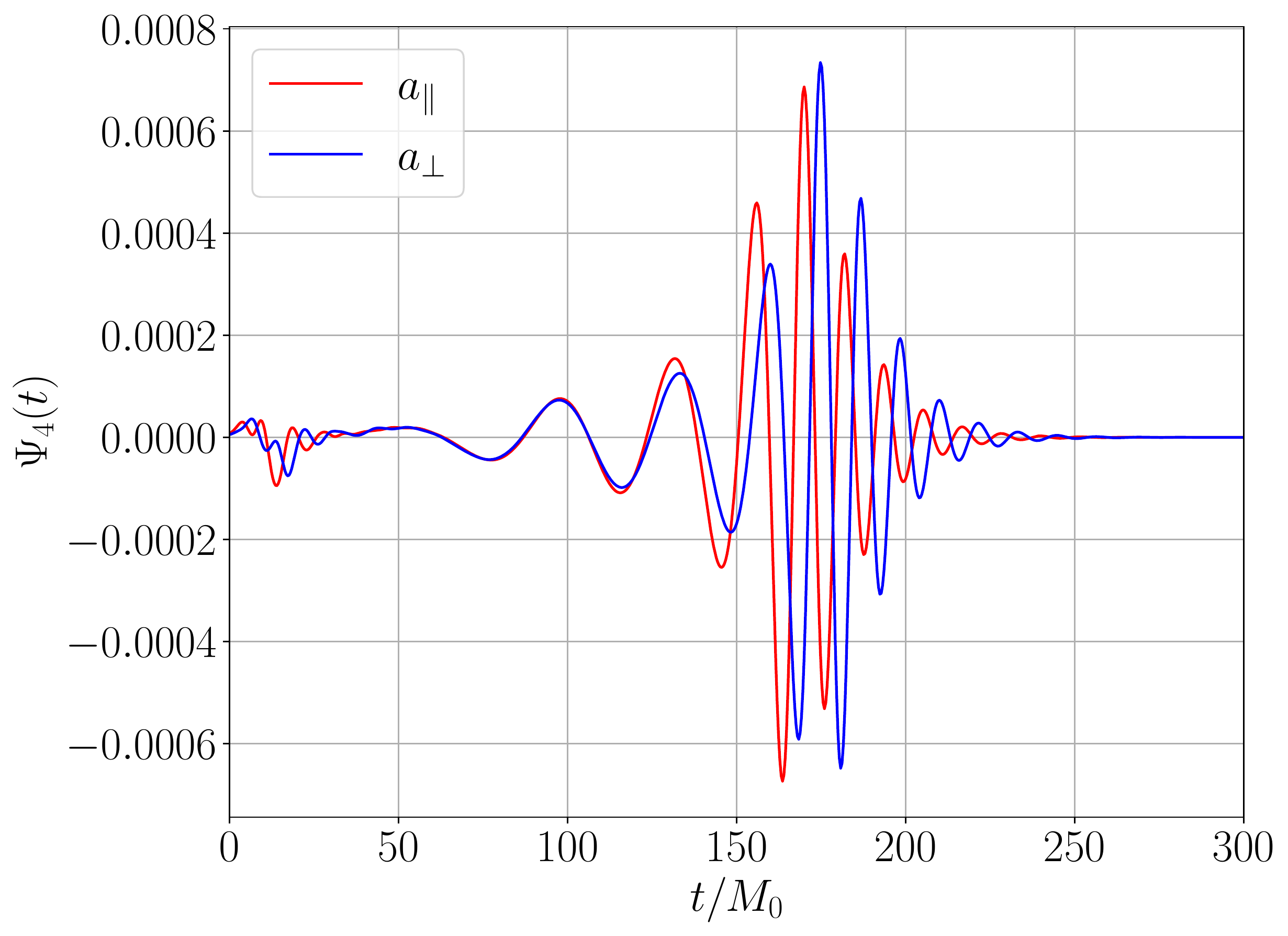}
    \includegraphics[width=0.325\linewidth]{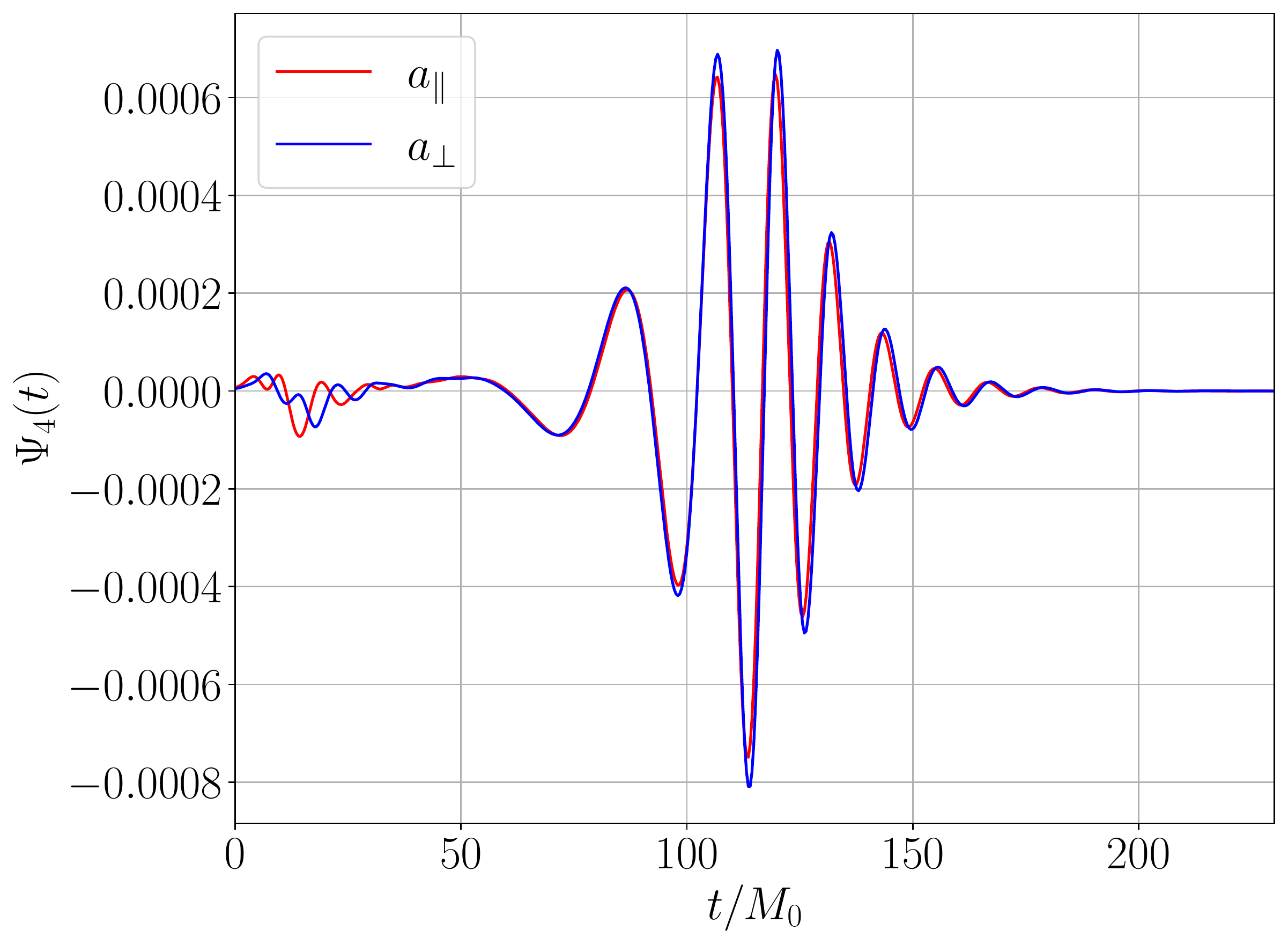}
	\caption{Mode $l=2$, $m=2$ of the Weyl scalar $\Psi_4$ for the equal mass, spinning \bh{} binaries. Panels  from left to right correspond to $\widehat\Pi_0= (5.0,~7.5,~10.0)$, respectively, with red lines for $a_\parallel$ and blue for $a_\perp$.}
	\label{fig:radn2s}
\end{figure*}

Figure~\ref{fig:radn2s} shows the mode $l=2$, $m=2$ of the Weyl scalar $\Psi_4$. Panels from left to right are for  $\widehat\Pi_0= (5.0,~7.5,~10.0)$, respectively, with red lines for $a_\parallel$ and blue for $a_\perp$. It is interesting to notice that for $\widehat\Pi_0 = 10.0$ there is very little difference in the (2,2) mode between the $a_\parallel$ and $a_\perp$ case, this in spite of the large difference they have, as we shall see, in kicks produced. After all, the $a_\perp$ cases are in the super-kick class. This means that the differences are in the higher modes. We also observe from the waveforms in Fig.~\ref{fig:radn2s} that, as for the un-equal mass and non-spinning \bh{} binaries, the larger the value of $\widehat\Pi_0$, the earlier the binary merges, and the reasons are similar. The accretion of scalar field by the \bh{} increases their masses and thus the luminosity of the binary.

\begin{figure*}[!htb]
	\includegraphics[width=0.525\linewidth]{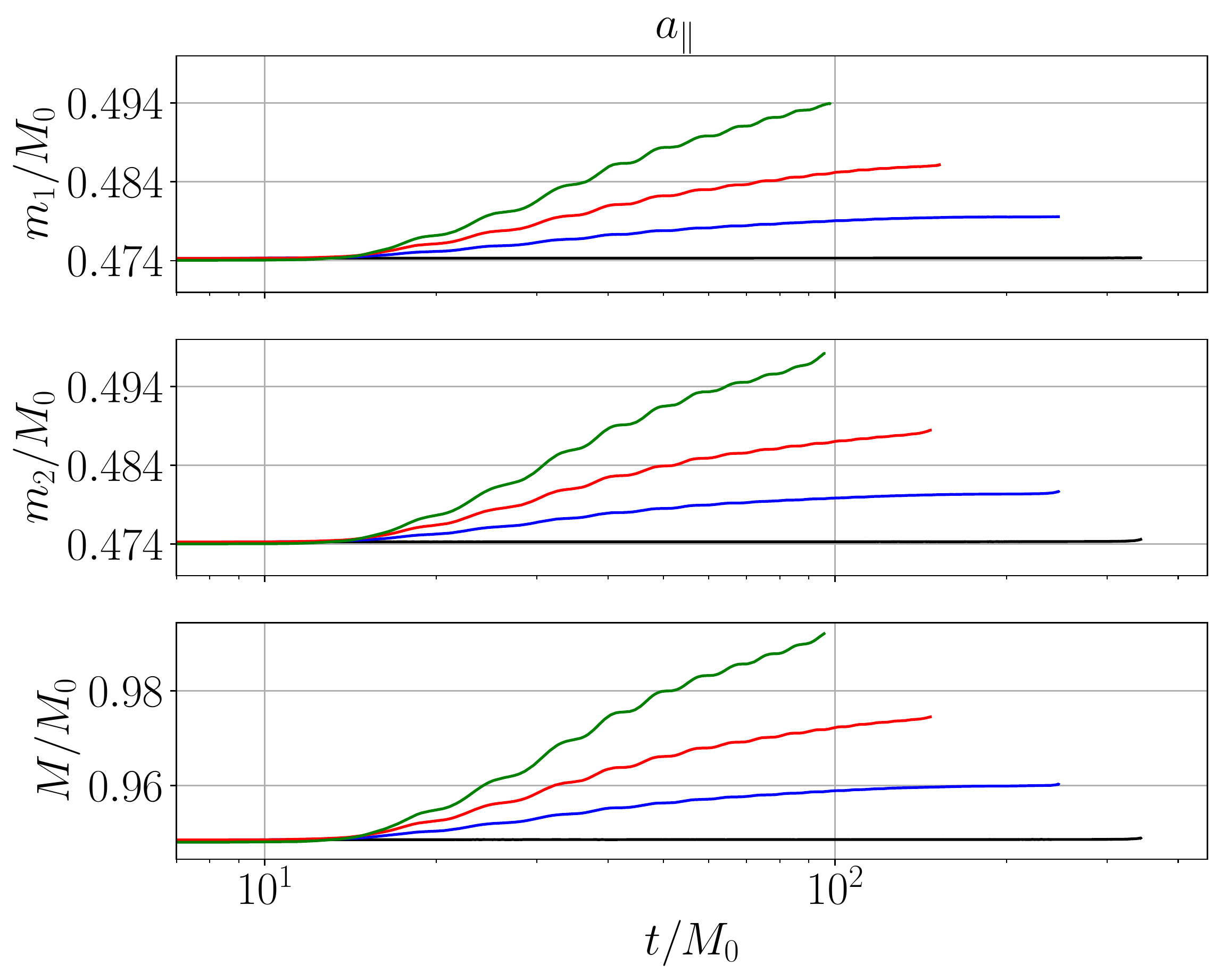}
	\includegraphics[width=0.464\linewidth]{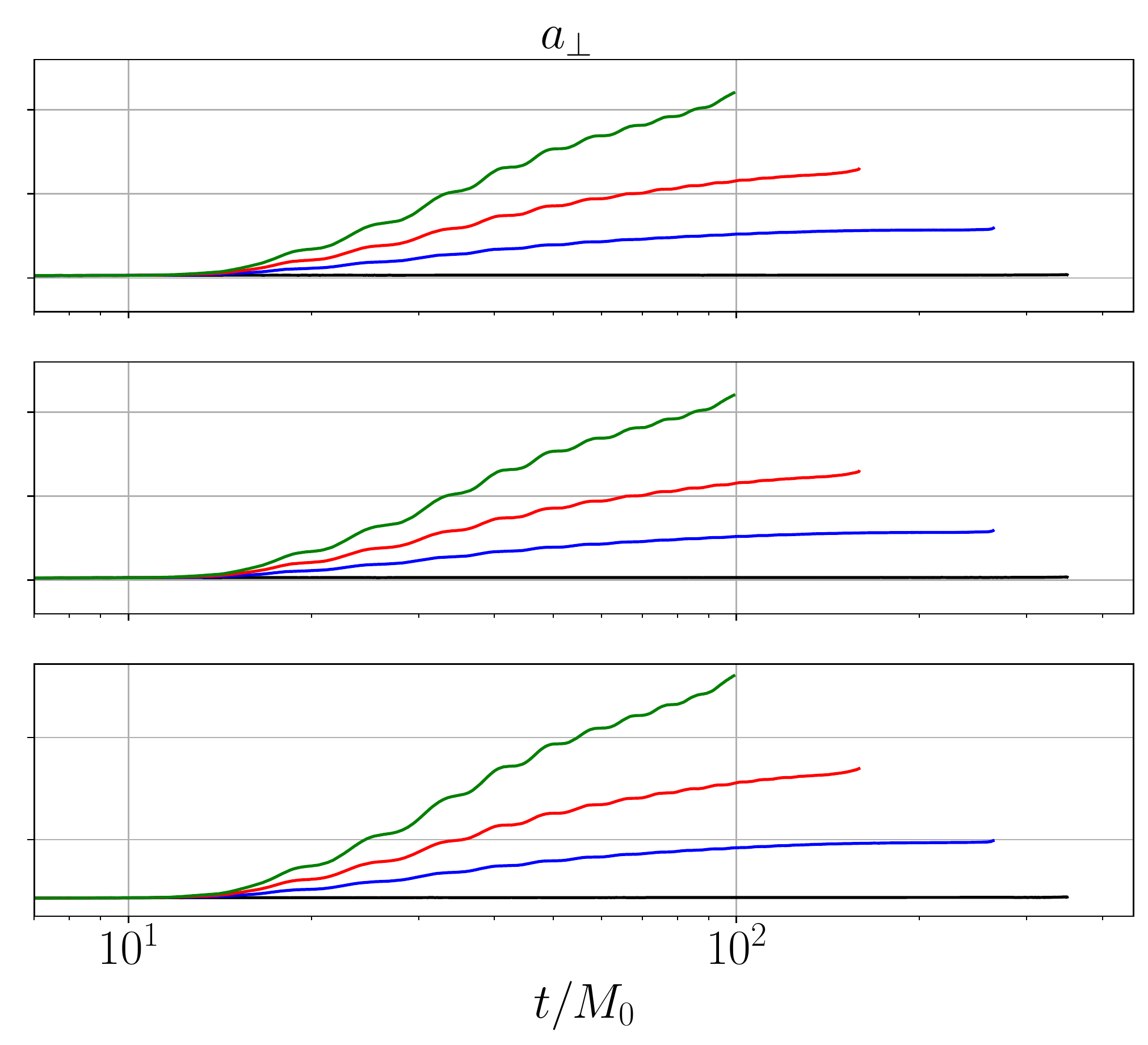}
	\caption{Evolution of $m_1$, $m_2$ and $M$ for equal mass, spinning \bh{}{} binaries cases. Left panel for  $a_\parallel$ and right panel $a_\perp$ (right). The colors black, blue, and green correspond to the scalar shell clouds with $\widehat\Pi_0=(0,~5.0,~7.5,~10.0)$ respectively. The lines end at the time when the merger occurs.}
	\label{fig:masses2}
\end{figure*}

Figure~\ref{fig:masses2}, shows from top to bottom the evolution of $m_1$, $m_2$, and $M$, respectively. Left panels are for the $a_\parallel$ cases and the right  ones for $a_\perp$. The line colors black, blue, red, and green correspond $\widehat\Pi_0=(0,~5.0,~7.5,~10.0)$, respectively. The behaviour in the growth of the masses is similar to that of unequal mass, non-spinning \bh{} binaries. Namely, the growth is monotonic with $\widehat\Pi_0$. Interesting to point out that the growth in $m_1$ and $m_2$ is identical in the $a_\perp$; thus, $q$ remains unity. This is because for both holes, the orientation of their spins relative to the orbital angular momentum, are identical. On the other hand, since for the $a_\parallel$ cases, the \bh{} with mass $m_1$ has its spin aligned with the orbital angular momentum and for the other anti-aligned, it is clear from panel top-left and middle-left that there is a slight difference in the growth between hole $m_1$ and $m_2$. The \bh{} with mass $m_2$ grows slightly more then $m_1$. This translates into mass ratios at merger of $q =1.0049,1.0073,1.0102$ for $\widehat\Pi_0 = 5.0,\, 7.5,\,10$, respectively. This is consistent with accretion of spinning black holes immersed in a gaseous environment or circumbinary disks~\cite{Lopez_Armengol_2021}.

\begin{figure*}[!htb]
	\includegraphics[width=0.32\linewidth]{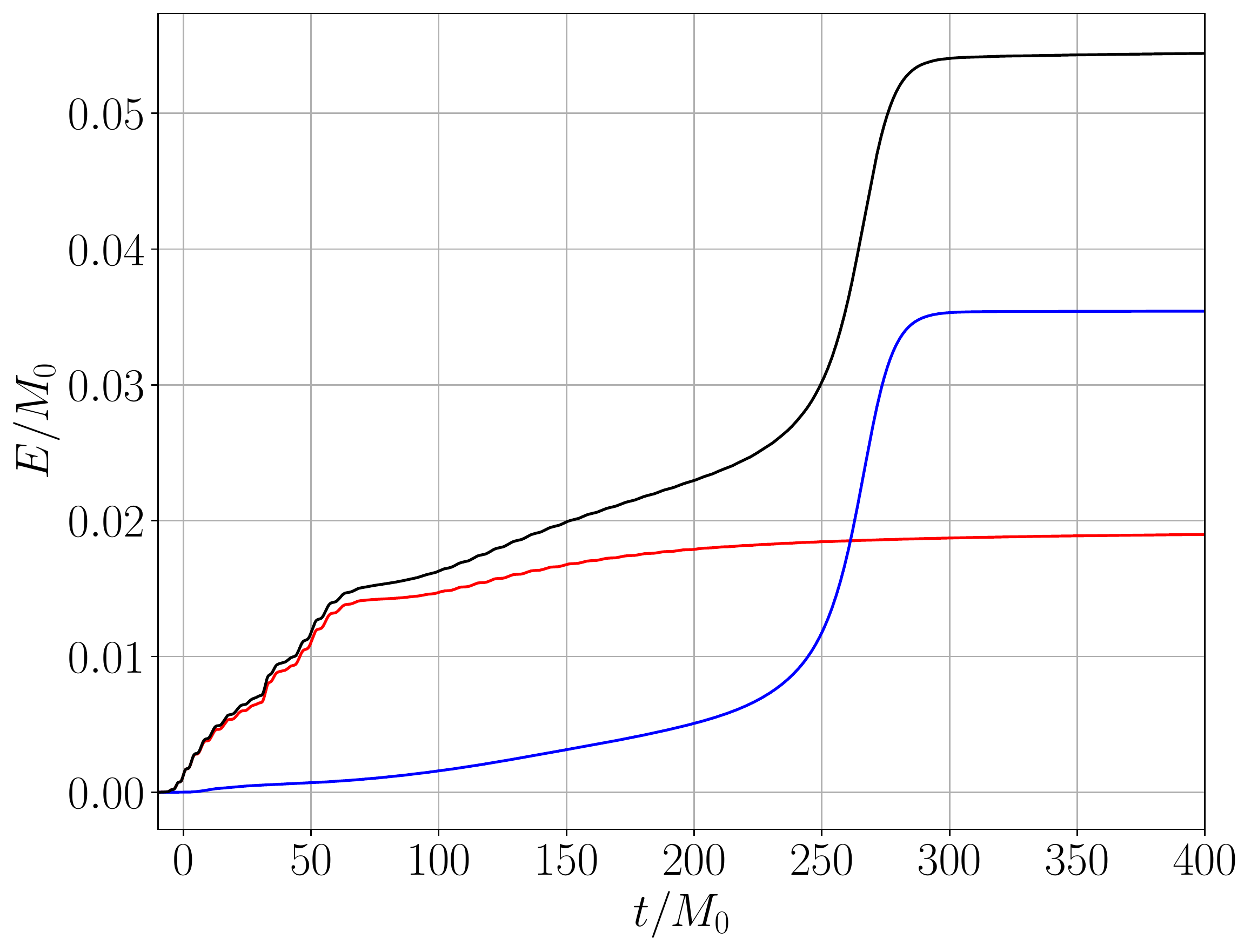}
	\includegraphics[width=0.32\linewidth]{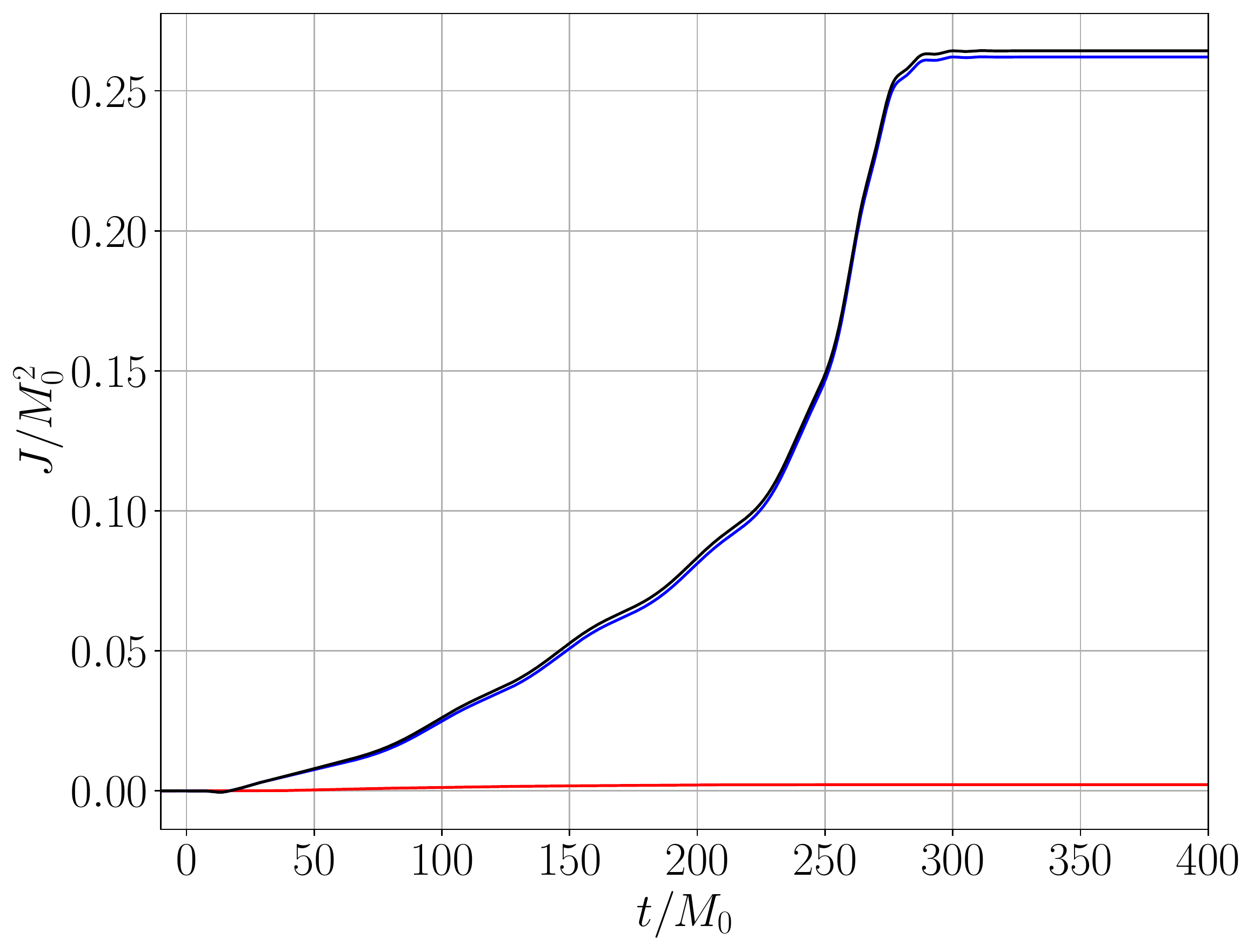}
    \includegraphics[width=0.33\linewidth]{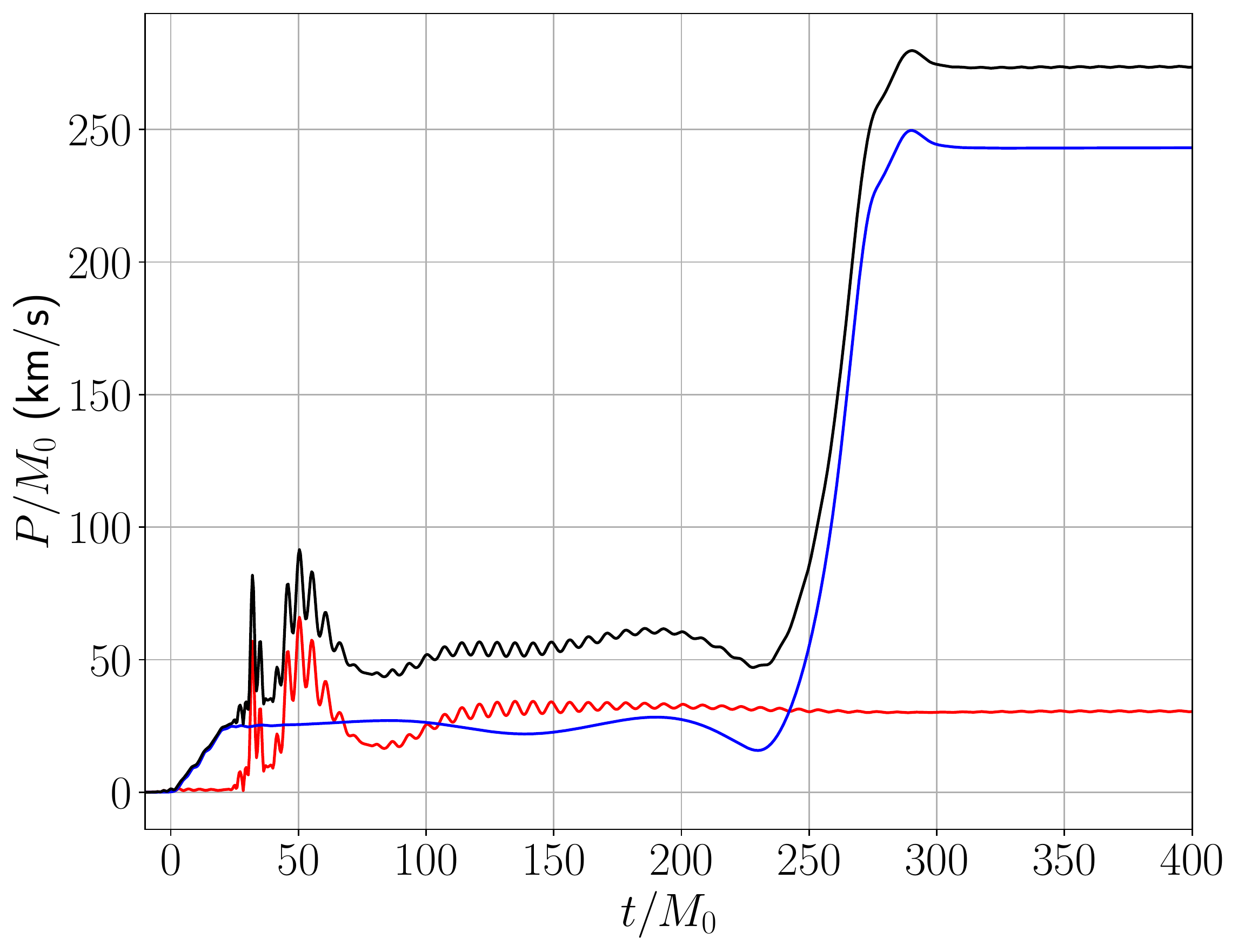}
	\includegraphics[width=0.32\linewidth]{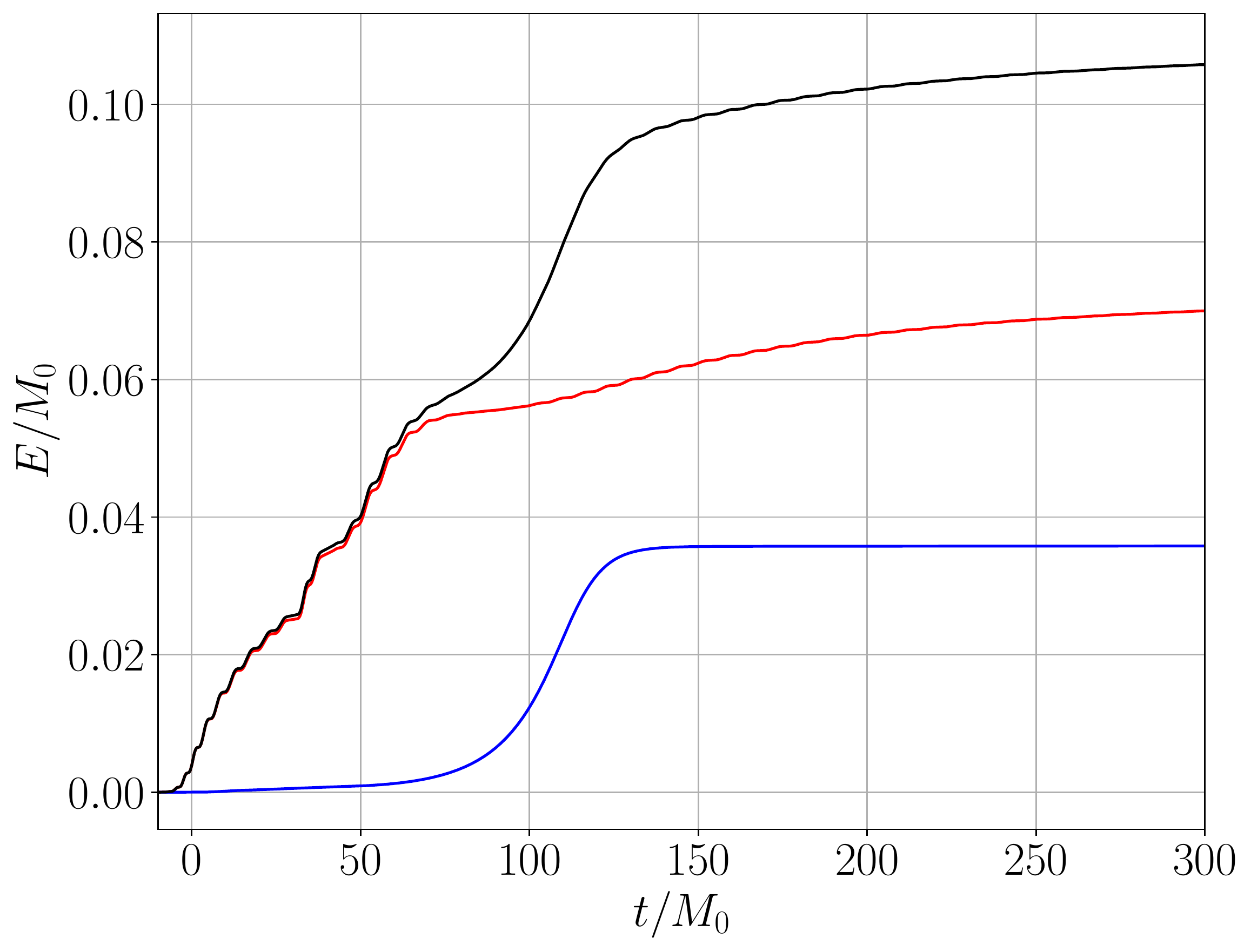}
	\includegraphics[width=0.32\linewidth]{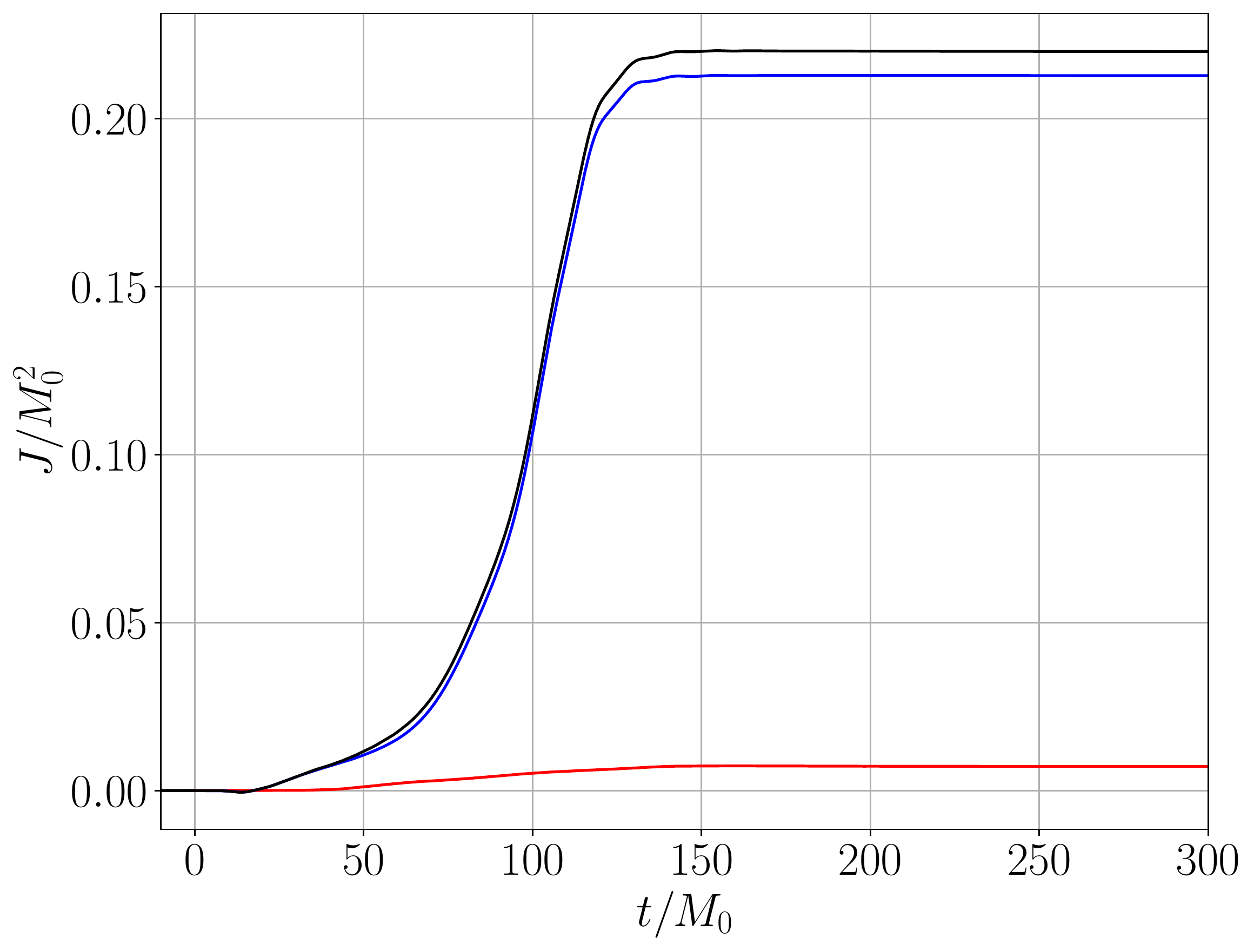}
    \includegraphics[width=0.33\linewidth]{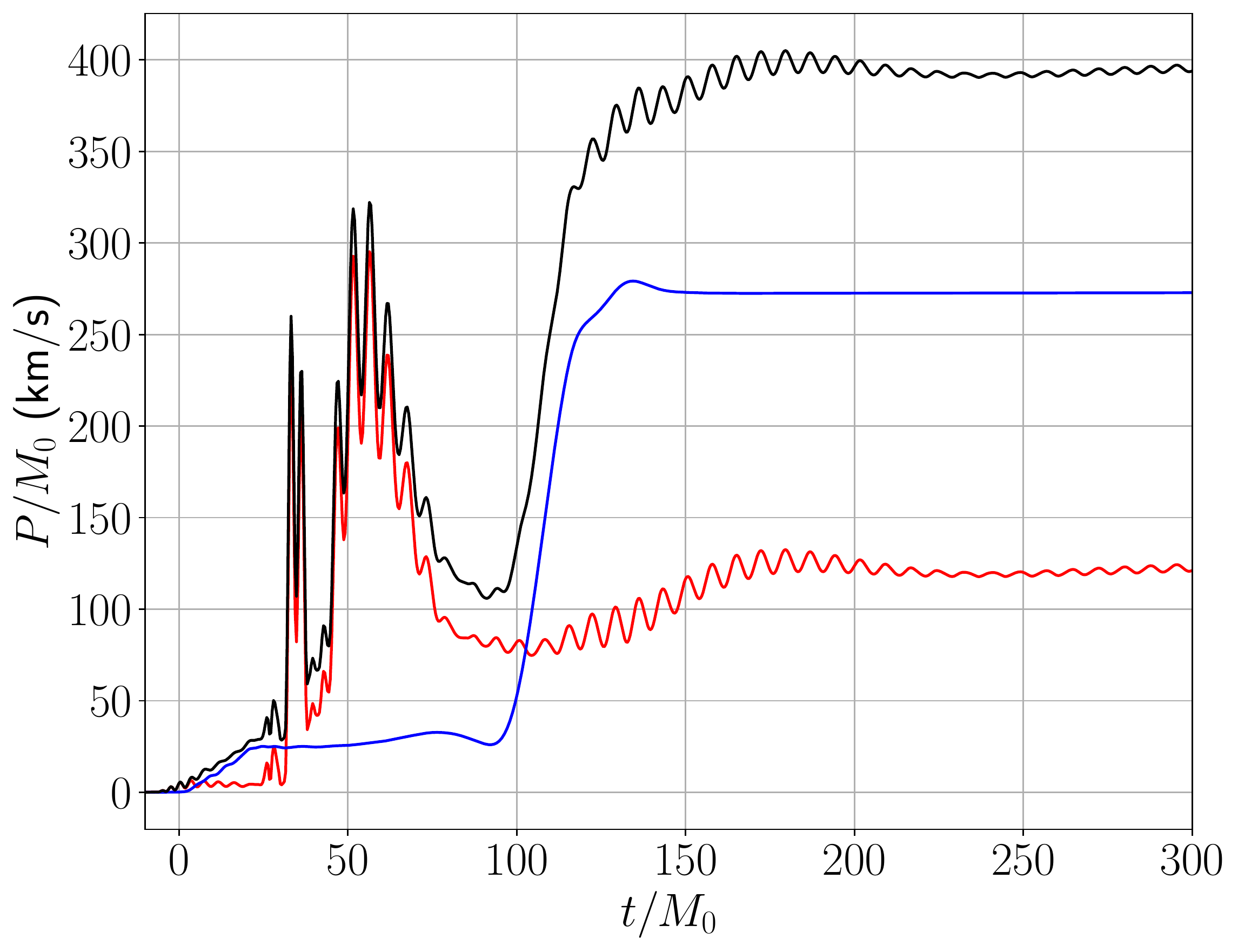}
	\caption{Energy (left panels), angular momentum (middle panels), and linear momentum (right panels) radiated as a function of time in \gw{s} (blue lines), in the scalar field (red lines), and the total (black lines) for the $a_\parallel$ cases,  with the top panels for $\widehat\Pi_0= 5.0$ and the bottom panels for $\widehat\Pi_0= 10.0$.}
	\label{fig:radn1s}
\end{figure*}	

Figure~\ref{fig:radn1s} shows the energy, angular momentum, and linear momentum radiated as a function of time in \gw{s} (blue lines), in the scalar field (red lines), and the total (black lines) for the $a_\parallel$ cases,  with the top panels for $\widehat\Pi_0= 5.0$ and the bottom panels for $\widehat\Pi_0= 10.0$. We observe that the angular and linear momentum radiated in \gw{s} is larger then in the scalar field for both $\widehat\Pi_0$ values. This is not the case for the energy radiated. Not surprisingly, the larger the value of $\widehat\Pi_0$, i.e. the larger the initial energy in the scalar field, the larger the energy emission. This does not imply that the remnant \bh{} will have a smaller mass. As we can see from Table~\ref{table:kick-velocity-table-spin} and saw from Fig.~\ref{fig:masses2}, the larger $\widehat\Pi_0$, the larger the final \bh{} because of the accretion of scalar field.

\begin{table}[!htb]
	\begin{center}
		\begin{tabular}{ c  c c c  }
			\hline
			\hline
			~Case~&$m_f/M_0$&$a_{f}$& $|v|$ (km/s) \\
			\hline
			$a_\parallel$000    &  ~0.9512~  &  ~0.6851~  &  302   \\
			$a_\parallel$050    &  ~0.9603~  &  ~0.6856~  &  285    \\
			$a_\parallel$075    &  ~0.9691~  &  ~0.6844~  &  297    \\
			$a_\parallel$100    &  ~0.9850~  &  ~0.6970~  &  362     \\
			\hline
			\hline
		\end{tabular}
	\end{center}
				\caption{Mass $m_f$, spin $a_f$ and magnitude of the kick of the final \bh{} for equal mass, spinning \bbh{s} in the $a_\parallel$ cases.}
							\label{table:kick-velocity-table-spin}
\end{table}

Regarding the radiated angular momentum from the middle panels of Fig.~\ref{fig:radn1s}, the scalar field emission is significantly smaller than  from \gw{s}. However, when comparing the emission in \gw{s} from $\widehat\Pi_0 = 5.0$ (top-middle panel) with that of $\widehat\Pi_0 = 10.0$ (bottom-middle panel), the former is slightly larger. Since the initial configuration has mostly orbital angular momentum because the spins are anti-aligned, this implies that the spin of the final \bh{} for $\widehat\Pi_0 = 5.0$ will be smaller than for $\widehat\Pi_0 = 10.0$, as we can see Table~\ref{table:kick-velocity-table-spin}. This is consistent because the $\widehat\Pi_0 = 10.0$ binary merges earlier (see Fig.~\ref{fig:radn2s}), and thus it does not radiate as much angular momentum as with the $\widehat\Pi_0 = 5.0$ case. 

The situation seems to reverse with the linear momentum radiated. Similar to the angular momentum radiated, it is still the case that, as the binary mergers earlier because of the presence of the scalar field, it does not ``accumulate" as much kick as in the vacuum case (see kick values for $a_\parallel$050 and $a_\parallel$075 in Table~\ref{table:kick-velocity-table-spin}). However, as we can see from the right panels in Figure~\ref{fig:radn1s}, the kick contribution from the scalar field increases with $\widehat\Pi_0$ and eventually turns things around. At $\widehat\Pi_0 = 10.0$  this contributions is such that the kick becomes larger than in the vacuum case. 

\begin{figure*}[!htb]
	\includegraphics[width=0.32\linewidth]{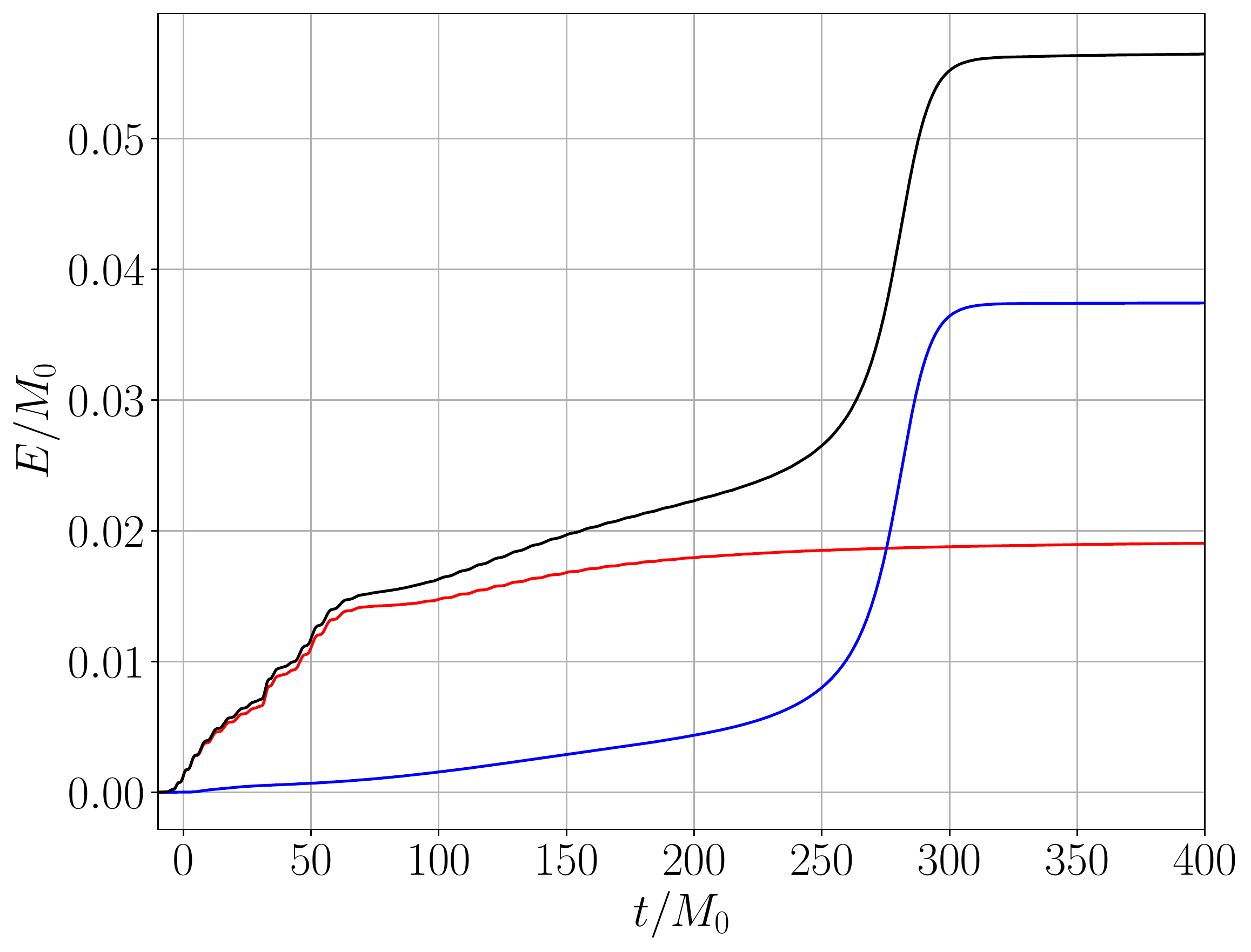}
	\includegraphics[width=0.32\linewidth]{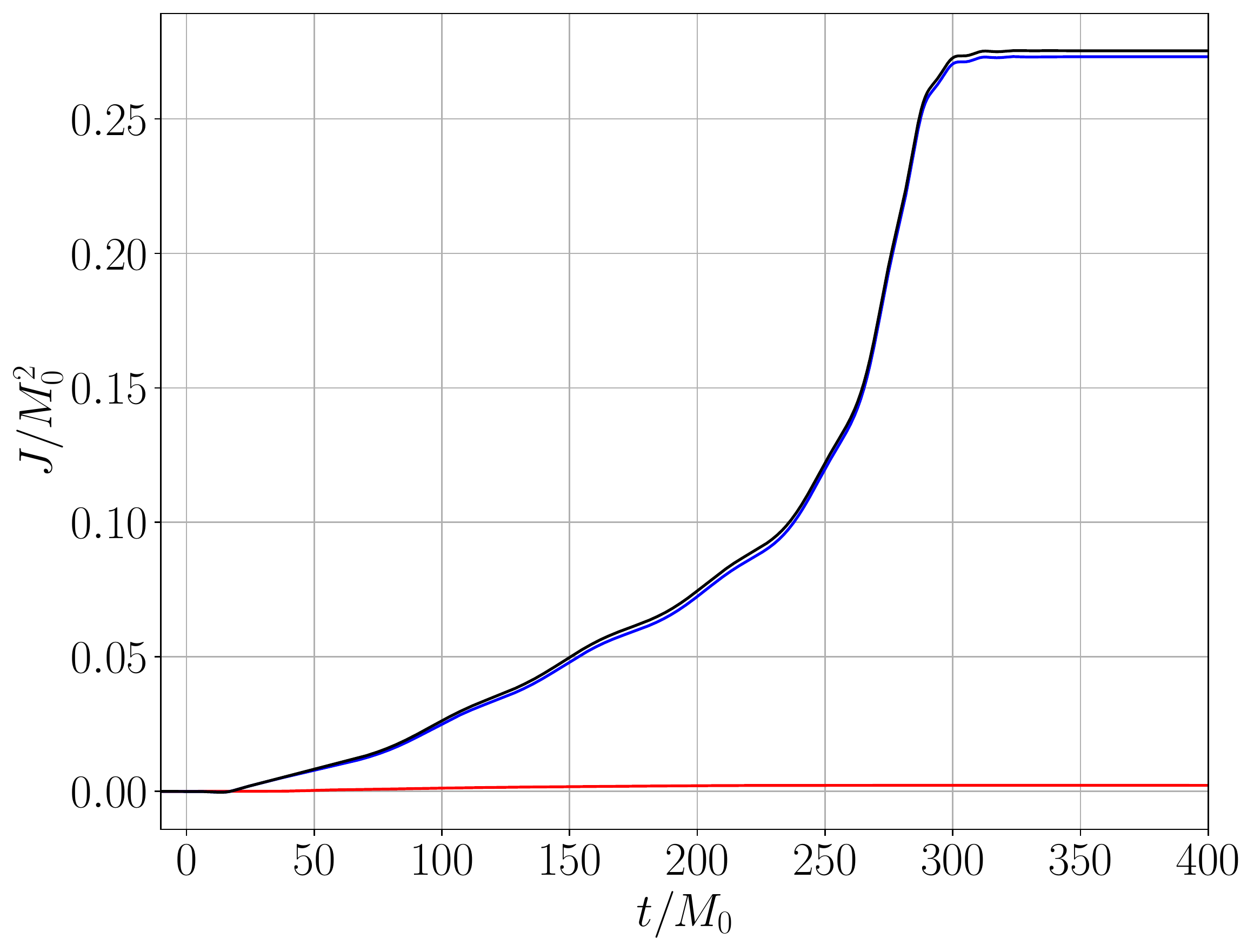}
    \includegraphics[width=0.33\linewidth]{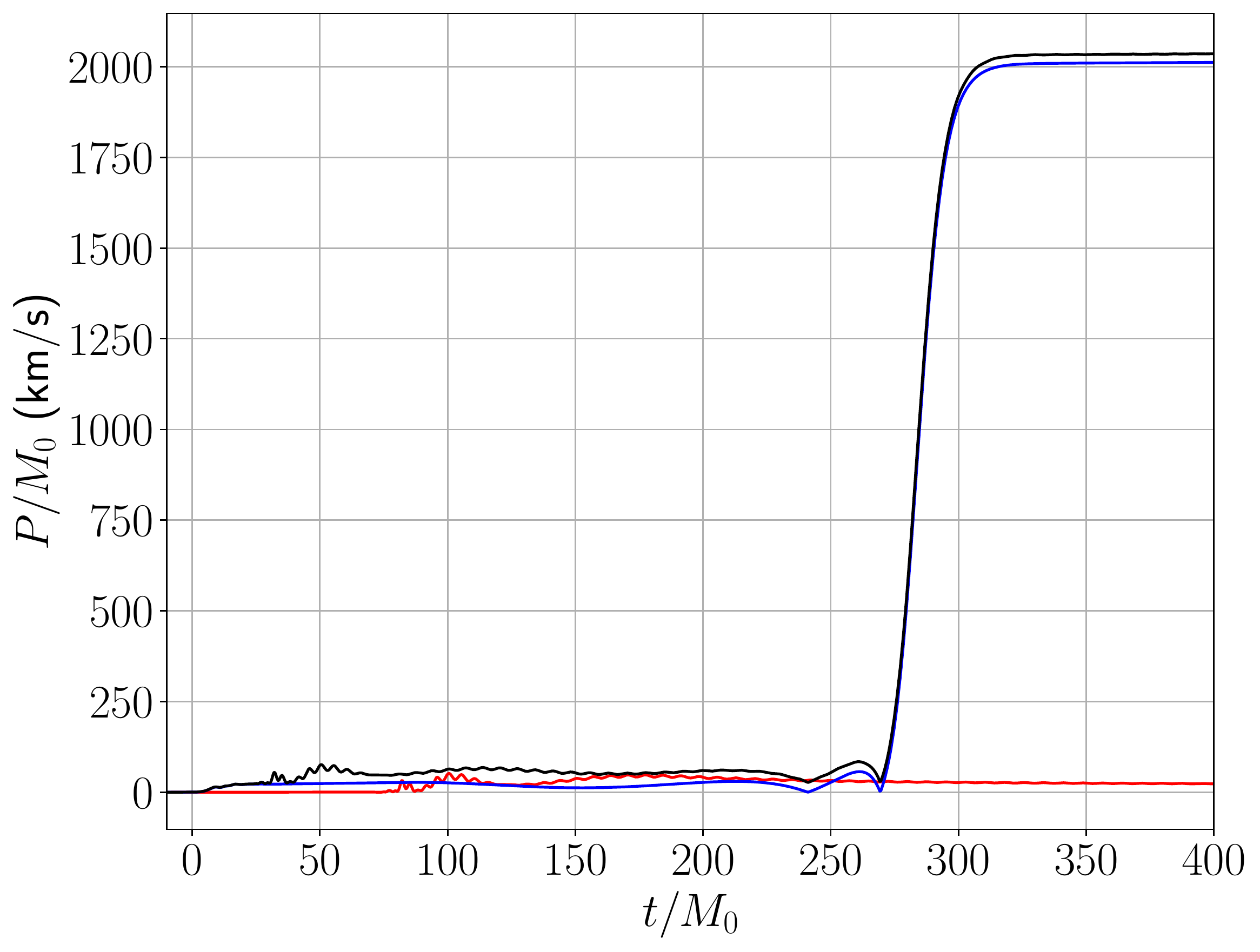}
	\includegraphics[width=0.32\linewidth]{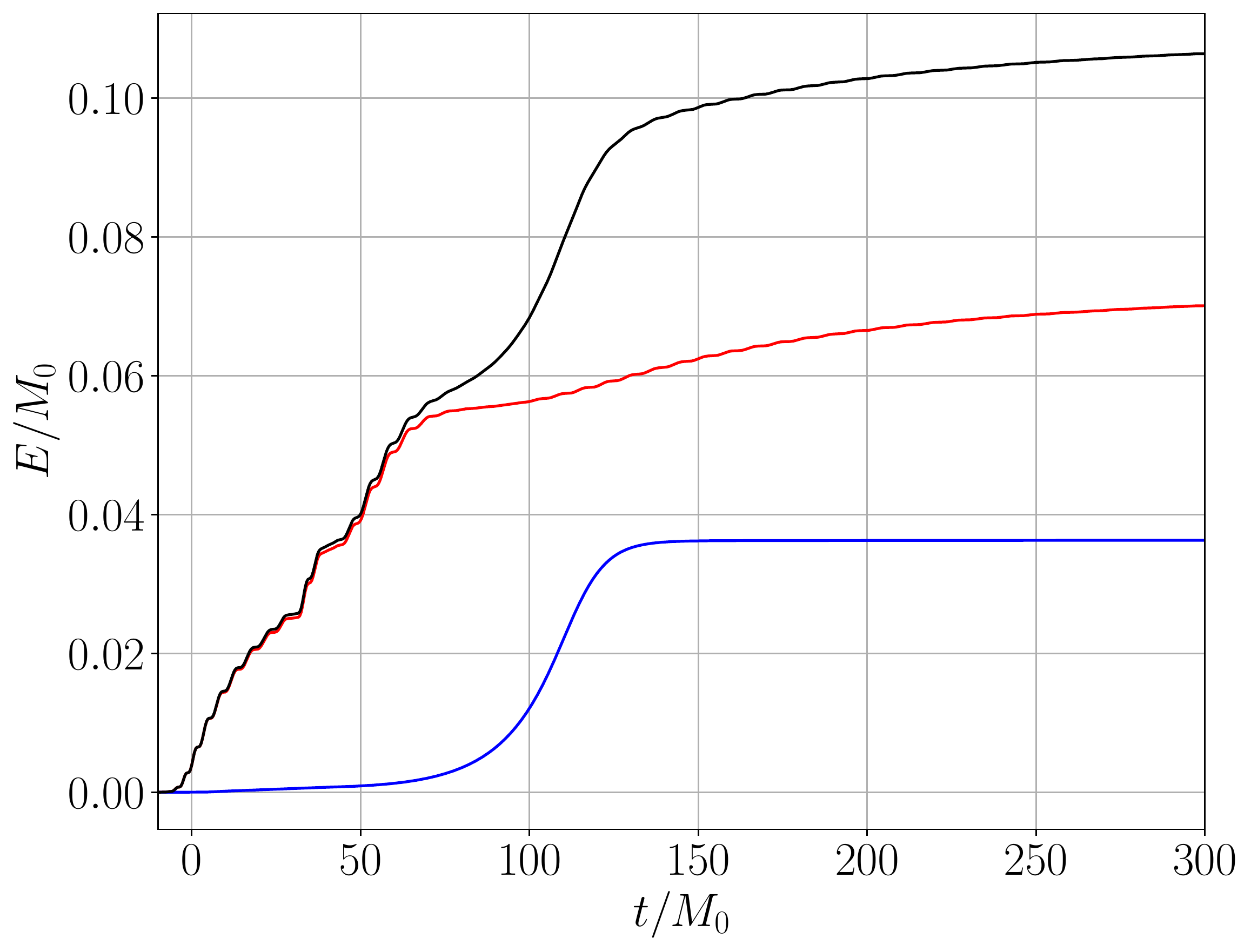}
	\includegraphics[width=0.32\linewidth]{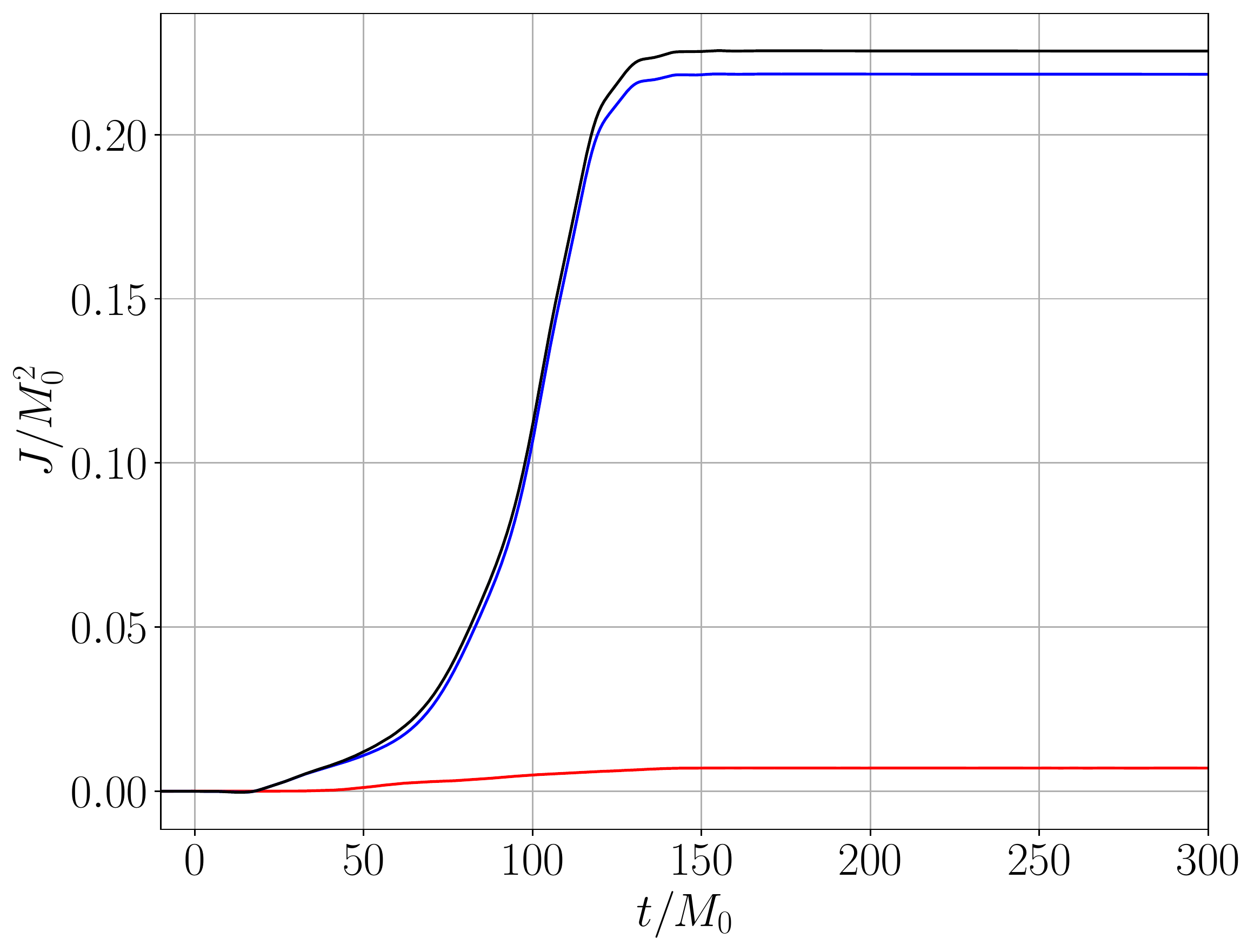}
    \includegraphics[width=0.33\linewidth]{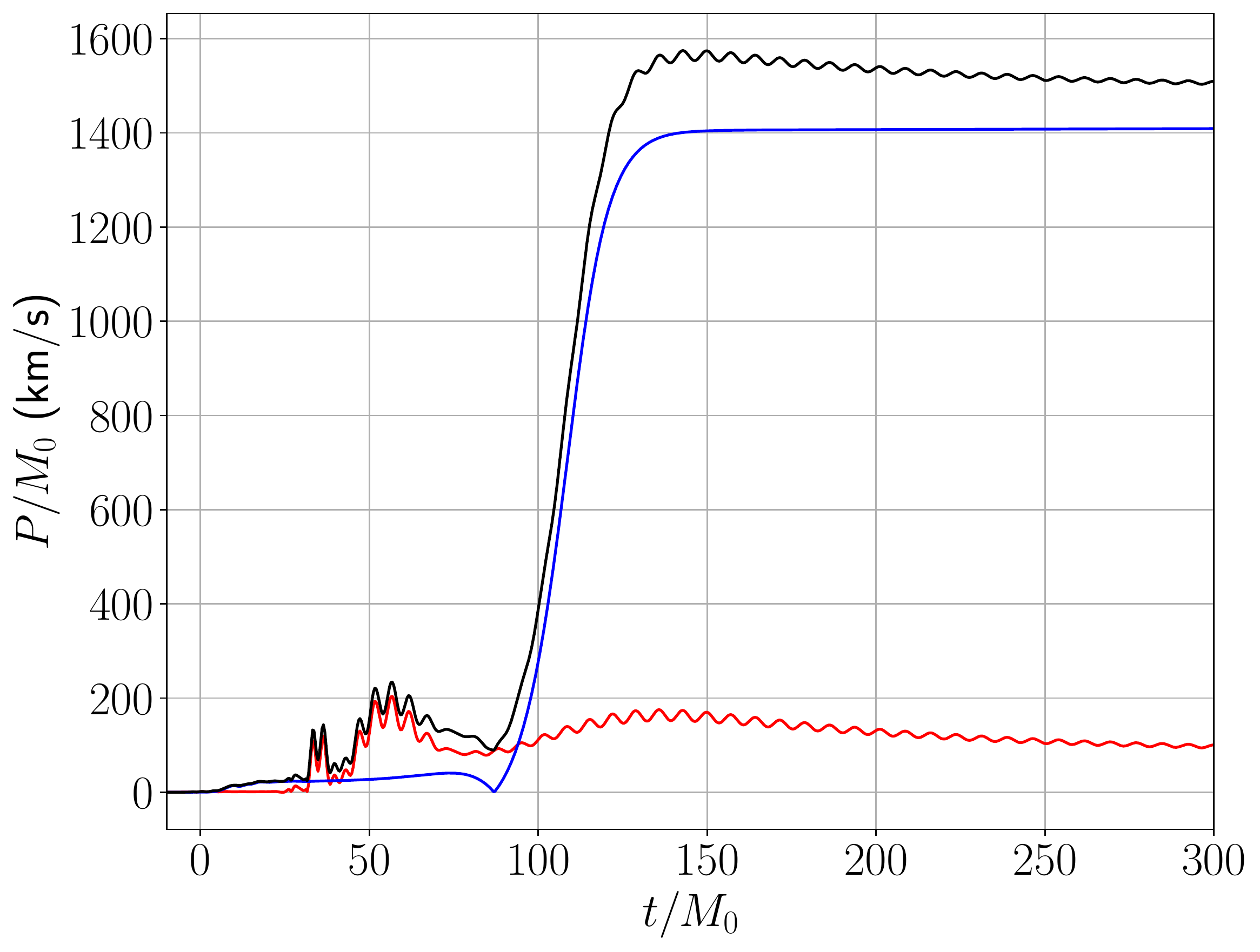}
	\caption{Energy (left panels), angular momentum (middle panels), and linear momentum (right panels) radiated as a function of time in \gw{s} (blue lines), in the scalar field (red lines), and the total (black lines) for the $a_\perp$ cases,  with the top panels for $\widehat\Pi_0= 5.0$ and the bottom panels for $\widehat\Pi_0= 10.0$.}
	\label{fig:radn1s2}
\end{figure*}

Figure~\ref{fig:radn1s2} shows the energy, angular momentum, and linear momentum radiated as a function of time in \gw{s} (blue lines), in the scalar field (red lines), and the total (black lines) for the $a_\perp$ cases,  with the top panels for $\widehat\Pi_0= 5.0$ and the bottom panels for $\widehat\Pi_0= 10.0$. It is interesting to observe that the spin configuration does not have a big effect on the energy and angular momentum radiated. Left and middle panels in Fig.~\ref{fig:radn1s2} are very similar to those in Fig.~\ref{fig:radn1s} for the $a_\parallel$ cases. The differences come in the linear momentum radiated (right panels in Fig.~\ref{fig:radn1s2}). Although the trend of which radiation dominates is similar to those in the $a_\parallel$ cases, the magnitude of the emission in the $a_\perp$ cases is much larger, after all these are super-kick setups.

Table~\ref{table:kick-velocity-table-spin2} shows the mass, spin, and the $z$-component kick velocity (the most dominant in this cases) for the $a_\perp$ cases. Regarding the mass of the final \bh{}, for the same reasons as all the previous binary types, $m_f$ increases monotonically with $\widehat\Pi_0$. There seems to be also monotonicity with $\widehat\Pi_0$ in $a_f$. The reason is because the larger the value of $\widehat\Pi_0$, the faster the binary merges thus the lower the angular momentum radiated and the larger residual angular momentum that goes into the final spin.   

\begin{figure}[!htb]
	\includegraphics[width=1.0\linewidth]{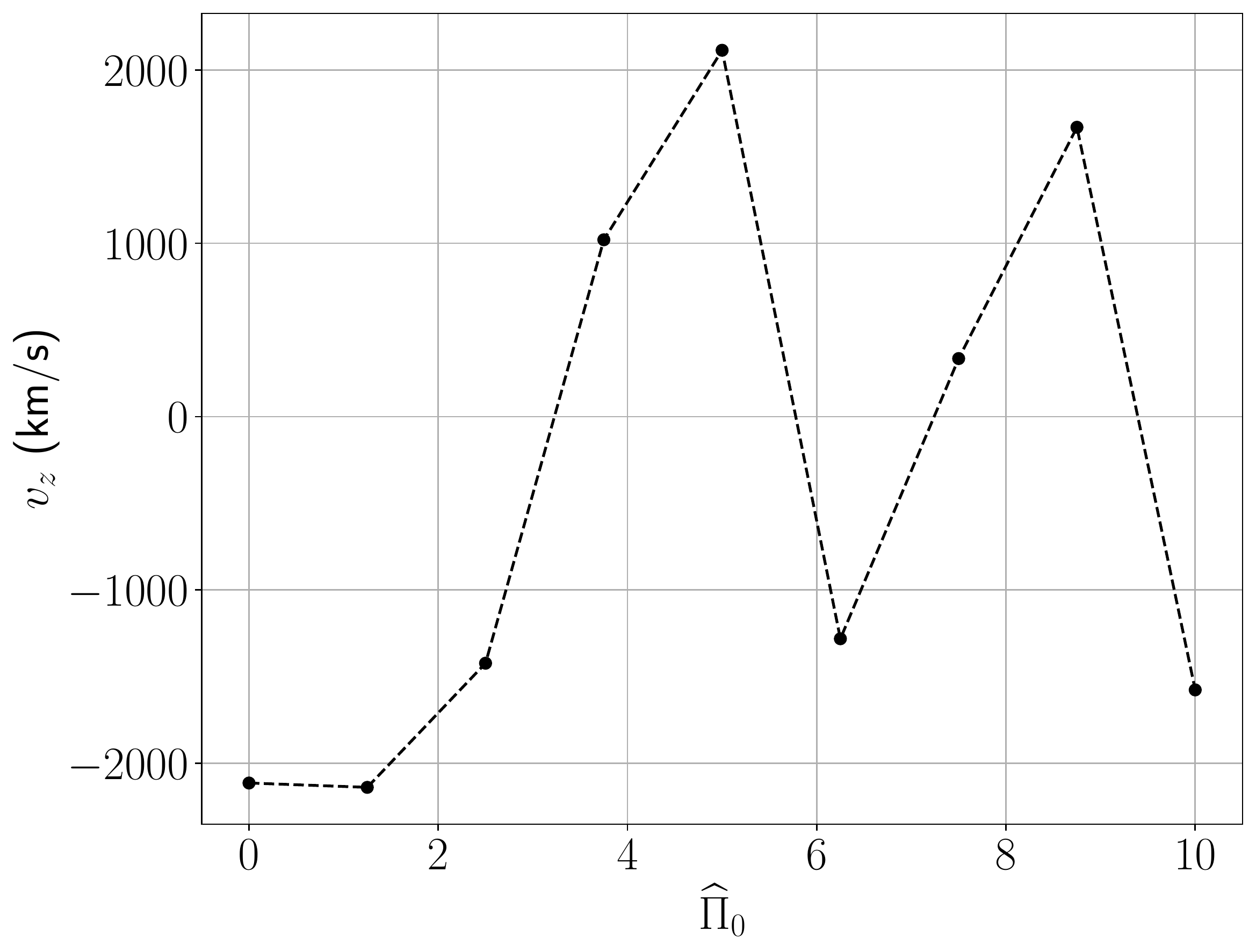}
	\caption{Final kick as a function of $\widehat\Pi_0$ for the $a_\perp$ (super-kick setup) values in Table~\ref{table:kick-velocity-table-spin2}}
	\label{spin_kicks}
\end{figure}

There is no monotonicity in the kicks. To help understand the situation, we plot the kicks as a function of $\widehat\Pi_0$ in Fig.~\ref{spin_kicks}. In this figure, we observe hints of an oscillatory trend in the $z$-component of the kick as a function of $\widehat\Pi_0$. The reason for this oscillatory behaviour is similar to the one found in the first studies of super-kicks, namely that the magnitude and direction of the kick is proportional to the cosine of the angle that the in-plane components of the spins make with the infall direction at merger~\cite{Campanelli_2007}. In the vacuum case, this dependence is obtained by changing the be initial direction of the spins. In our case, it is the effect that the scalar field has on the mass growth of the holes, and thus its orbital dynamics, that produces the changes of the spin alignment relative to the infall direction.

\begin{table}[!htb]
	\begin{center}
		\begin{tabular}{ c  c c c  }
			\hline
			\hline
			~Case~&$m_f/M_0$&$a_{f}$& $v_z$ (km/s) \\
			\hline
			$a_\perp$0000   &  ~0.9500~  &  ~0.6797~  &  -2113     \\
			$a_\perp$0125   &  ~0.9500~  &  ~0.6786~  &  -2138       \\
			$a_\perp$0250   &  ~0.9515~  &  ~0.6801~  &  -1422      \\
			$a_\perp$0375   &  ~0.9560~  &  ~0.6860~  &  1020      \\
			$a_\perp$0500   &  ~0.9582~  &  ~0.6802~  &  2113     \\
			$a_\perp$0625   &  ~0.9650~  &  ~0.6834~  &  -1281     \\
			$a_\perp$0750   &  ~0.9691~  &  ~0.6829~  &  335     \\
			$a_\perp$0875   &  ~0.9734~  &  ~0.6848~  &  1669      \\
			$a_\perp$1000   &  ~0.9841~  &  ~0.6966~  &  -1576  \\

			\hline
			\hline
		\end{tabular}
	\end{center}
				\caption{Mass $m_f$, spin $a_f$ and $z$-component of the kick of the final \bh{} for equal mass, spinning \bbh{s} in the $a_\perp$ cases}
							\label{table:kick-velocity-table-spin2}
\end{table}

\section{Conclusions}\label{sec:conclusions}

We have presented results from a numerical study of \bbh{} mergers immersed in a scalar field cloud, focusing on the effects that the cloud has on the gravitational recoil, as well as on the spin of the final \bh{.} We considered two initial configuration scenarios: binaries with non-spinning, un-equal mass \bh{s} and binaries with equal mass \bh{s} and their spins anti-aligned spins. For the later case we had to subcategories, one in which the \bh{} spins were parallel to the orbital angular momentum (i.e. non-precessing), and the other with the \bh{} spins in the orbital plane in the so-called super-kick setup. The initial geometry of the scalar field cloud was a thin shell encapsulating the binary. 

In all cases, because of scalar field accretion, the \bh{s} gained mass, thus increased the emission of \gw{s}, and as a consequence accelerated the merger. This also induced changes in the mass ratio of the binary, with the exception of the binaries in the super-kick configuration because the spin relative to the orbital momentum were the same.

We computed the radiated energy, angular momentum, and linear momentum emitted in both the \gw{} and scalar field channels. For the un-equal mass \bh{} binaries, we found that the scalar field emission was dominant in energy and linear momentum. Because of the later, the kicks were larger than in the vacuum case. Since the emission of angular momentum by the scalar field was smaller than from \gw{s}, the spins varied very little from their vacuum counterparts. A similar situation took place with the equal mass, spinning \bh{} binaries; the presence of the scalar field did not translate into significant changes in the spin of the final \bh{} relative to the vacuum case. The main reason for this general situation is because the initial scalar field cloud did not have angular momentum that could be transferred via accretion to the \bh{s}.

Regarding the gravitational recoil of the final \bh{}, for the case of unequal mass, non-spinning \bh{} binaries, we obtained that 
the kicks were larger that their vacuum counterparts because in these configurations the emission of linear momentum is larger via the scalar field channel. Some of the kicks reached super-kick levels of $\sim 1,300$ km/s. 

For the binaries with equal mass \bh{s} and spins aligned with the orbital angular momentum, we observed two effects competing against each other as we increased $\widehat\Pi_0$. The scalar field accretion increased the \bh{} masses and accelerated the merger. This ameliorated the ``accumulation" of the kick. Acting in the opposite direction was the linear momentum radiated in the scalar field, increasing with the value of $\widehat\Pi_0$ and eventually yielding kicks larger than in the vacuum case. 

Finally, for equal mass and spinning \bh{} binaries in the super-kick configuration, we observed hints of the oscillatory behavior observed in the vacuum case. The reasons are similar; that is, the magnitude and direction of the kick is proportional to the cosine of the angle that the in-plane components of the spins make with the infall direction at merger~\cite{Campanelli_2007}. However, instead of this dependence be from changing the initial direction of the spins, in our case, it is the change in the dynamics of the binary from the mass growth of the holes that produces the changes of the spin alignment relative to the infall direction. 

One, of course, must take our results with a grain of salt regarding astrophysical implications. The purpose of our study was solely to investigate the sensitivity of \bbh{} merger dynamics and the resulting final \bh{} to the presence of a scalar field. Our results should be taken as a guide of the scale of energy in a scalar field necessary to imprint noticeably effects on the merger time of the binary and the gravitational recoil of the final black hole. In a subsequent study, we will focus on the impact in parameter estimation under the eyepiece of \gw{} analysis.

\section{Acknowledgments}

This work was supported in part by NSF awards 2207780 and 2114582 to PL and DS, the National Key Research and Development Program of China (Grant No. 2020YFC2201503), the National Natural Science Foundation of China (Grants No. 12105126, No. 11875151, No. 11705070, No. 12075103, and No. 12047501), the China Postdoctoral Science Foundation (Grant No. 2021M701531), the 111 Project under (Grant No. B20063), and the Fundamental Research Funds for the Central Universities (Grant No. lzujbky-2021-pd08). 

\section*{References}
\bibliographystyle{unsrt}

\end{document}